\documentclass{ws-rv961x669}
\usepackage{ws-rv-van}     % numbered citation/references (default)
\usepackage{ws-rv-thm}     % comment this line when `amsthm / theorem / ntheorem` package is used
\usepackage{subfigure}     % required only when side-by-side / subfigures are used
\usepackage{hyperref}
\hypersetup{
    colorlinks,
    citecolor=blue,
    filecolor=black,
    linkcolor=blue,
    urlcolor=blue
}
\usepackage{color}

\def\kms{\ifmmode{~{\rm km~s^{-1}}}\else{~km s$^{-1}$}\fi}
\def\cc{\ifmmode{~{\rm cm^{-3}}}\else{~cm$^{-3}$}\fi}
\def\fesc{\ifmmode{{f_{esc}}}\else{$f_{\rm esc}$}\fi}
\def\fstar{\ifmmode{{f_\star}}\else{$f_\star$}\fi}
\def\lsim{\lower0.3em\hbox{$\,\buildrel <\over\sim\,$}}
\def\gsim{\lower0.3em\hbox{$\,\buildrel >\over\sim\,$}}

\def\cubecm{\ifmmode{~{\rm cm^{-3}}}\else{cm$^{-3}$}\fi}
\def\Ms{\ifmmode{~{\rm M_\odot}}\else{M$_\odot$}\fi}
\def\Zs{\ifmmode{~{\rm Z_\odot}}\else{Z$_\odot$}\fi}
\def\h2{H$_2$}

\def\hh{H$_2$}

\newcommand{\Msigma}{M-$\sigma$}

\newcommand\unit[1]{\; \textrm{#1}}
\newcommand\ion[2]{#1$\;${\small\rmfamily\@Roman{#2}}\relax}

\newcommand{\new}[1]{{\color{black}#1}}

\makeindex
\begin{document}

\chapter[The formation of the first black holes]{The formation of the first black holes\label{ch9}}

\author[Wise]{John H. Wise}%\footnote{Author footnote.}}
%\index[aindx]{Author, F.} % or \aindx{Author, F.}
%\index[aindx]{Author, S.} % or \aindx{Author, S.}

\address{Center for Relativistic Astrophysics, School of Physics, \\
Georgia Institute of Technology, Atlanta, GA 30332, USA \\
jwise@physics.gatech.edu}

\begin{abstract}
    The most massive black holes at redshifts $z = 6$ were already over billion solar masses.  In this chapter, we discuss the formation and growth of the first black holes in the Universe.  The deaths of massive primordial stars provide potential seeds of supermassive black holes.  Theoretical models predict that the seed black hole masses range from 10 to 100,000 solar masses.  Their initial fueling may be limited by feedback from its progenitor star, the black hole itself, and nearby star formation.  Once the halo and galaxy surpasses a critical mass, black hole growth may accelerate as the central gravitational potential deepens with strong ensuing star formation.
\end{abstract}

%\markboth{Even Page Header}{Odd Page Header} % Customized running heads

\body

%\tableofcontents

\section{Introduction}
\label{ch9:sec:intro}

% 2-3 pages
% a. Broad overview and motivation
% b. Importance of studying the formation and early growth of BHs
% c. Observational and theoretical implications, referring to other chapters

\begin{figure}
    \centering
    \includegraphics[width=\textwidth]{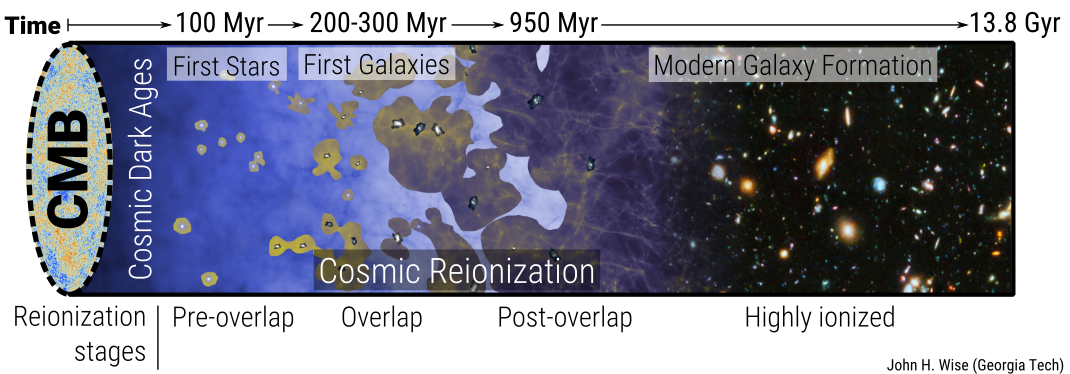}
    \caption{\label{fig:timeline} Cosmic timeline of the universe after recombination and the emission of the cosmic microwave background along with the stages of cosmic reionization. The galaxy survey image is taken from the {\it Hubble Ultra Deep Field}.}
\end{figure}

This chapter explores our current theoretical understanding of the formation and growth of the first black holes in the Universe.  The dawn of the galaxy formation era is marked by the formation of the first generations of stars from primordial gas within dark matter (DM) halos with masses of only $10^{5-8} \Ms$ \cite{Bromm13}.  These ``minihalos'' are the basic building blocks from which all galaxies assemble.  Within a billion years after the Big Bang, galaxies and their central black holes (BHs) grow from the merger of minihalos and young and small galaxies as the universe is reionized.  Figure \ref{fig:timeline} shows the approximate timing and sequence of galaxy formation and reionization.  These are the galaxies that exist at the frontier of detection for the Hubble Space Telescope (HST), but whose deeper origins will be probed by the James Webb Space Telescope (JWST).

\subsection{Observations of early supermassive black holes}

During Cosmic Dawn (redshifts $z \gsim 6$), observations of nearly 2{,}000 galaxies at $z = 6-10$ in the HST legacy fields have shown that galaxies are growing rapidly from pre-galactic gas\cite{Oesch16, Bouwens22}, but observational constraints are fewer for their central BHs\cite{Yang21}.  Beyond this observational frontier, theoretical models and numerical simulations of early structure formation guide our expectations for future electromagnetic and gravitational wave observations of these galaxies and their BHs.  Successful predictions must match the overall statistics of SMBH growth histories while accounting for the extremes.  For instance, growth rates in periodic growth events can explain more typical SMBHs in low-redshift galaxies, whereas this more typical behavior is challenged by the $\sim 200$ behemoth SMBHs with masses often exceeding $10^9 \Ms$ at $z \ge 6$\cite{Wu15,Yang21}.

Observations of the most distant and luminous quasars during Cosmic Dawn only probe the tip of the iceberg.  For example, the most distant observed quasar at $z = 7.54$ is powered by a SMBH with a mass of $\sim 8 \times 10^8 \Ms$ \cite{Banados18}.  These extreme objects are however rare with a comoving number density of about one per cubic gigaparsec\cite{Fan06}. Cosmic structure forms hierarchically with numerous smaller objects merging together to form rare massive halos.  Thus, there should be many more low luminosity and inactive supermassive black holes (SMBHs) with moderate growth histories in more normal galaxies like our Milky Way\cite{Gultekin09,McGreer18}.

The present-day correlation between the SMBH mass and the stellar velocity dispersion, known as the \Msigma{} relation, suggests that there is some connection between galaxy and BH growth but can be difficult to determine observationally at high redshift.  It is not guaranteed nor expected that early BHs and galaxies begin on a relation, but they must approach the observed \Msigma{} relation as the population evolves toward the present-day.

For the interested reader, Chapter 6 gives a more detailed account of the current observational status of SMBHs.  One can combine the \Msigma{} relation and the galaxy luminosity (or halo mass) function to ascertain that the total BH mass density in SMBHs with masses above $10^6 \Ms$ rises from $\sim 10^{-3} \Ms$ per comoving Mpc$^{3}$ at $z \sim 7$ to the present-day value of $5 \times 10^5 \Ms$ per comoving Mpc$^{3}$.\cite{Somerville08, Willott10, Volonteri17}.  These individual and statistical measures provide critical high-redshift constraints on any early BH growth model.

\subsection{Overview of supermassive black hole seeding}

Even more mysterious and unknown is the origin of SMBHs.  The formation mechanisms, often called seeding, is currently under debate\cite{Woods19}.  The mainstream scenarios involve the BH remnants of massive metal-free stars, whose initial mass function is thought to favor massive star formation but is only weakly constrained\cite{Bromm13,Prole22}.  SMBH formation models can be categorized into {\it light} and {\it heavy} seeds that are approximately delineated at $1000 \Ms$ and range between $10 \Ms$ and $10^5 \Ms$.  Important questions that revolve around BH seeding concern their formation environment, their initial mass function, their growth, and their connection to their host galaxies and DM halos, all of which this chapter will address.

%   We first review primordial star formation in Section \ref{ch9:sec:pop3} and then discuss different BH seeding mechanisms in Section \ref{ch9:sec:seeds}.  We then turn our attention to their subsequent growth, or lack of, within the first generations of galaxies in Section \ref{ch9:sec:grow}.  We close this chapter with a summary and a discussion of outstanding questions to be answered with the next generation of space- and ground-based telescopes and gravitational wave detectors.

\section{Primordial star formation}
\label{ch9:sec:pop3}

% 6 pages
% a. Metal-free chemistry, lack of cooling
% b. Simulations: stellar masses, host halos, redshifts, feedback
% c. Stellar multiplicity: binaries, dense clusters
% d. Top-heavy IMF? Implications on black hole formation and compact object binaries
% e. Supermassive star formation: refer back to Heger+ chapter and focus on the conditions in which they might occur

The first generations of stars form from primordial gas that does not have any elements heavier than lithium, i.e. metals, because it has not been processed by stellar fusion and supernovae since the Big Bang.  Astronomers have historically classified stars by their metallicity.
\begin{description}
    \item[Population I:] metallicities similar to our Sun that has 1.3\% metals by mass \cite{Asplund09};
    \item[Population II:] metallicities less than 1/10$^{\rm th}$ of the solar value.
\end{description}
The term Population III (Pop III) was used for metal-free stars as early as 1967 by Partridge \& Peebles\cite{Patridge67}.  A truly metal-free star has never been observed to-date.  In principle, they should have existed at early times.  If a Pop III star had a mass less than $0.8 \Ms$, its lifetime is longer than the current age of the Universe, and it should still exist today.  They may be the hidden needles in the haystack of billions of old stars within present-day galaxies\cite{Tumlinson10}.  There are several observational campaigns to search for extremely metal-poor and even metal-free stars within the Milky Way and nearby dwarf galaxies.  The most metal-poor star discovered to-date, SDSS J102915+172927, only has a total metallicity of $10^{-4.3}$ of solar metallicity\cite{Caffau11}.

Most theoretical models suggest that Pop III stars are generally massive with typical masses in the tens of solar masses.\cite{Hirano15,Hosokawa16}  A significant fraction will thus leave behind BH remnants, providing a large source of potential SMBH seeds.

\subsection{Primordial gas cooling}

The star formation process begins by the gas cooling and losing its thermal pressure so that it can gravitationally collapse.  In present-day star formation, \new{dust grains, atomic electronic transitions in metals and molecular rotational and vibrational transitions, in particular thermal dust grain emission and CO line emission}, provide the bulk of the cooling of molecular clouds.  However without metals, primordial gas cooling primarily relies on \hh{} cooling in the gas-phase, where free electrons act as a catalyst in the following reactions
\begin{equation}
\begin{aligned}
    \textrm{H} + \textrm{e}^- &\rightarrow \textrm{H}^- + \gamma\\
    \textrm{H}^- + \textrm{H} &\rightarrow \textrm{H}_2 + \textrm{e}^-
\end{aligned}
\end{equation}
A similar set of reactions with deuterium and the HD molecule becomes important when electron fractions are enhanced by ionizing radiation from nearby star formation.  When the universe recombines and becomes nearly neutral 380{,}000 years after the Big Bang, there still exists a residual free electron fraction on the order of $10^{-5}$, which is insufficient to efficiently cool primordial gas.  

\subsection{Dark matter halos}

The basic building block of large-scale structure is the DM halo.  Density perturbations will decouple from the expansion of the universe and gravitationally collapse.  The collapsing DM comes into virial equilibrium, where its kinetic energy is twice the gravitational potential energy.  In spherical symmetry at high redshift (i.e. when the cosmological constant is unimportant), the resulting object is a sphere with an average overdensity of $\Delta_{\rm c} = 18\pi^2$ relative to the critical density of the universe $\rho_{\rm c} = (1+z)^3 \rho_{\rm c,0}$ at redshift $z$.  The present-day critical density $\rho_{\rm c,0} = 3H_0^2/8\pi G$ delineates forever expanding (open) and eventually collapsing (closed) universes and enters many cosmology calculations.  Here $H_0$ is the Hubble constant that describes the current expansion of the universe.

A halo with a virial mass $M_{\rm vir}$ and an average density $\rho_{\rm h} = \Delta_{\rm c} \rho_{\rm c}$ gives the virial radius
\begin{equation}
\begin{aligned}
    r_{\rm vir} &= \left( \frac{3 M_{\rm vir}}{4\pi \rho_{\rm h}} \right)^{1/3} = \left[ \frac{ 2 G M_{\rm vir}}{\Delta_{\rm c} H_0^2 (1+z)^3} \right]^{1/3}\\
    &= 1.00 \left( \frac{M_{\rm vir}}{10^8 \Ms} \right)^{1/3} \left( \frac{\Delta_{\rm c}}{18\pi^2} \right)^{-1/3} \left( \frac{h}{0.7} \right)^{-2/3} \left( \frac{1+z}{10} \right)^{-1} \unit{kpc}.
\end{aligned}
\end{equation}
The halo has a corresponding circular velocity
\begin{equation}
    \begin{aligned}
    V_{\rm c} &= \left( \frac{GM_{\rm vir}}{r_{\rm vir}} \right)^{1/2}\\
    &= 20.8 \left( \frac{M_{\rm vir}}{10^8 \Ms} \right)^{1/3} \left( \frac{\Delta_{\rm c}}{18\pi^2} \right)^{1/6} \left( \frac{h}{0.7} \right)^{1/3} \left( \frac{1+z}{10} \right)^{1/2} \unit{km s}^{-1}
    \end{aligned}
\end{equation}
from which we can define a virial temperature by equating the ``thermal energy'' of the DM to its kinetic energy
\begin{equation}
    \label{eqn:tvir}
    \begin{aligned}
    T_{\rm vir} &= \frac{\mu m_{\rm H} V_{\rm c}^2}{2k_{\rm b}}\\
    &= 1.57 \times 10^4 \left( \frac{M_{\rm vir}}{10^8 \Ms} \right)^{2/3} \left( \frac{\Delta_{\rm c}}{18\pi^2} \right)^{1/3} \left( \frac{h}{0.7} \right)^{2/3} \left( \frac{1+z}{10} \right) \left( \frac{\mu}{0.6} \right) \unit{K},
    \end{aligned}
\end{equation}
where $\mu$, $m_{\rm p}$, $k_{\rm b}$ are the mean molecular weight, proton mass, and Boltzmann constant, respectively.  When gas falls into a DM halo, it shock-heats to around the virial temperature.  At $T_{\rm vir} \sim 10^3\unit{K}$ or equivalently $M_{\rm vir} \sim 10^6 \Ms$, hydrogen is partially ionized to a free electron fraction on the order of $10^{-4} - 10^{-3}$.  Primordial gas with this ionization state can efficiently cool through \hh{} formation.  Once the gas can efficiently cool, it may become gravitationally unstable to collapse and subsequent star formation.  On the other hand, \hh{} can be easily destroyed by distant radiation sources.

\subsection{Molecular hydrogen dissociation: Lyman-Werner radiation}

\begin{figure}[t]
    \centering
    \includegraphics[width=\textwidth]{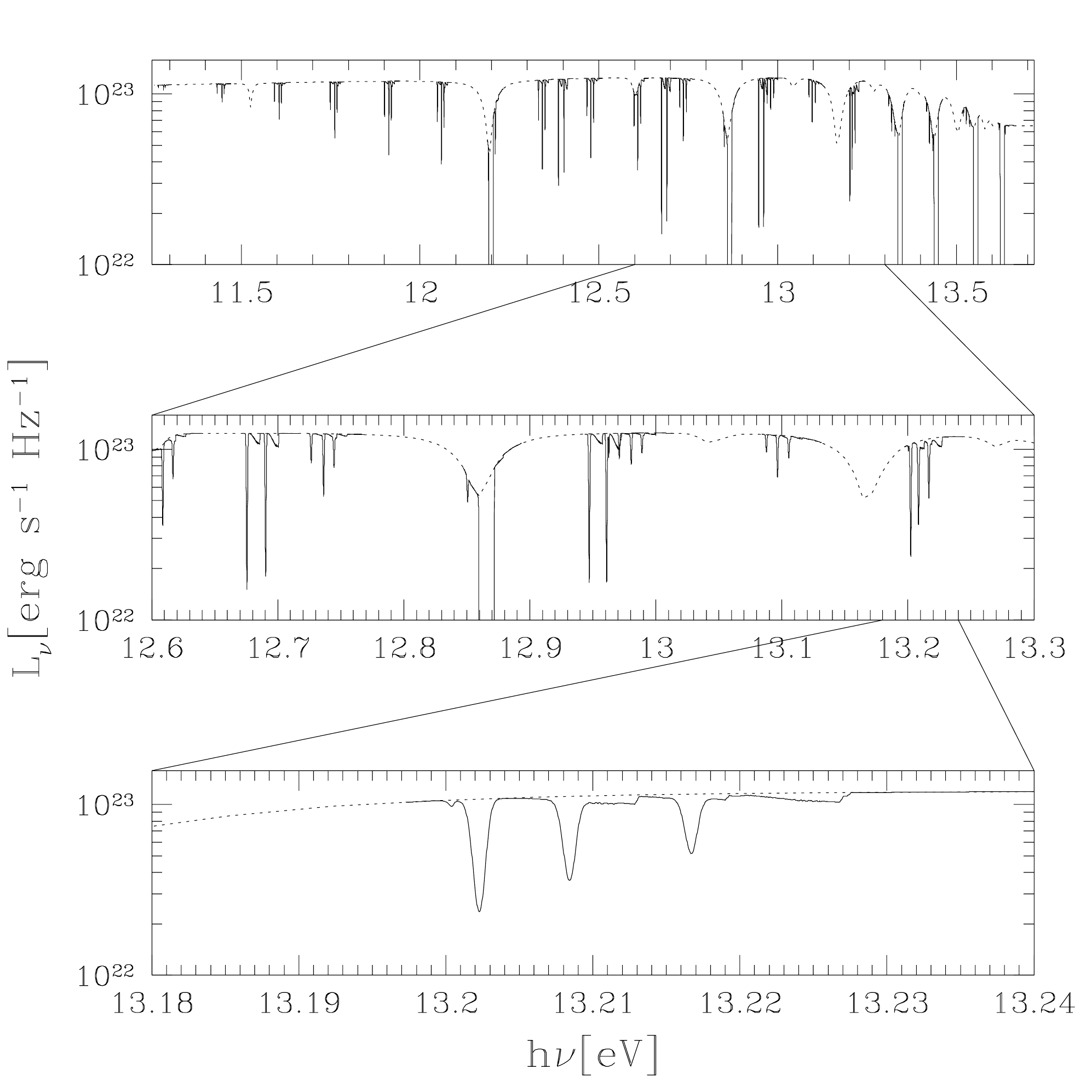}
    \caption{\label{fig:lw}\textbf{Lyman-Werner bands.} Absorption spectrum of a Pop III star only considering \hh{} photo-dissociation from the Lyman-Werner bands and the Lyman lines of atomic hydrogen.  Adapted from Ricotti {\it et al.}\cite{Ricotti01}}
\end{figure}

Molecular hydrogen is a fragile molecule that can be dissociated in the Lyman-Werner (LW) bands that is composed of 76 lines between 11.1 and 13.6~eV, shown in Figure \ref{fig:lw}.  Dissociation occurs through the Solomon process\cite{Field66, Stecher67} that is a two-step process, where first \hh{} is rotationally or vibrationally excited by a photon in the LW band
\begin{equation}
    \text{H}_2 + \gamma \rightarrow \text{H}_2^*.
\end{equation}
Next this excited \hh{} molecule has a probability of dissociating into two hydrogen atoms
\begin{equation}
    \text{H}_2^* \rightarrow \text{H} + \text{H}.
\end{equation}
Additionally, \hh{} formation can be weakly suppressed through infrared photons\cite{Wolcott12} at 0.76~eV that photo-detach the extra electron of the intermediary product H$^-$.

Being below the ionization potential of hydrogen of 13.6~eV, the universe is mostly optically thin to radiation in the LW bands.  This allows for the cosmological population of the first generations of stars and galaxies to build up a LW radiation background.  The only substantial sources of opacity are the Lyman transitions (de-excitation from level $n$ to the ground state) of hydrogen that exist between 10.2 and 13.6~eV at $E_n(\text{Lyman}) = 13.6 (1 - 1/n^2) \unit{eV}$.\cite{Haiman97}.  Molecular gas can also self-shield itself from the LW background, where the cloud's envelope absorbs radiation protecting the interior \hh{}.  The transmission fraction $f_{\rm sh}$ of the incoming radiation can be approximated with the power law
\begin{equation}
    f_{\rm sh} = \min\left[ 1, \left( \frac{N_{\rm H2}}{10^{14} \unit{cm}^{-2}} \right)^{-3/4} \right]
\end{equation}
in a system with a \hh{} column density $N_{\rm H2}$.  More accurate functional fits account for the thermal broadening and Doppler shifting of the absorption lines and different numerical approximations to the column density.\cite{Draine96,Wolcott11,Wolcott19}.

\subsection{Host dark matter halos}
\label{sec:halos}

\subsubsection{Population III stars}

Pop III stars form in the centers of DM halos.  They need to be massive enough to partially ionize hydrogen to catalyze \hh{} formation that happens above $\sim 2 \times 10^5 \Ms$\cite{Haiman96, Tegmark97}.  If a moderate LW radiation background exists, more massive halos are required to host increased \hh{} formation rates to negate the dissociation of \hh{}.  The critical halo mass to host Pop III star formation in a LW background is well approximated\cite{Visbal14} by
\begin{equation}
    M_{\rm crit} = 2.5 \times 10^5 \left( \frac{1+z}{26} \right)^{-1.5} \left[ 1 + 6.96(4\pi J_{\rm JW})^{0.47} \right] \Ms,
\end{equation}
and matches with cosmological simulations of Pop III star formation.  Here $J_{\rm LW}$ is the LW background intensity in units of $10^{-21}$ erg s$^{-1}$ cm$^{-2}$ Hz$^{-1}$ sr$^{-1}$. \new{In addition, baryons have a relative streaming velocity $v_{\rm bc}(\mathbf{x})$ with a Gaussian distribution of $\sigma_{\rm rms} \approx 30\kms$ at recombination relative to dark matter.\cite{Tselia10}.  This difference is caused by baryons being coupled to radiation in the early universe, whereas dark matter is not coupled.  It fluctuates on scales larger than a few comoving Mpc and can suppress gas condensation and star formation in minihalos\cite{Tselia11, Naoz11}.  The effects of Lyman-Werner radiation and streaming velocities interact non-linearly\citep{Kulkarni21, Schauer21}.  The critical halo mass including both processes is well approximated\cite{Schauer21} by
\begin{equation}
    \log (M_{\rm crit} / \Ms) = 5.562 \times (1 + 0.279 J_{21}^{0.5}) + s (v_{\rm bc}/\sigma_{\rm rms})
\end{equation}
with $s = 0.614 + (1 - 0.560 J_{21}^{0.5})$.}  Lastly, these more massive halos are more likely to self-shield their centers and may form in halos with masses less than $M_{\rm crit}$ as a reservoir of \hh{} accumulates\cite{Skinner20, Kulkarni21}.

A secondary heating source is the virial heating of the halo as it grows, i.e. the increase in $T_{\rm vir}$.  This dynamical heating can manifest itself in shocks in a turbulent medium that can add both thermal and turbulent pressures, preventing the center from gravitationally collapsing\cite{Yoshida03}.  The dynamical heating rate is calculated by
\begin{equation}
    \Gamma_{\rm dyn} \equiv \frac{\text{d}T_{\rm vir}}{\text{d}t} = \alpha M_{\rm vir}^{-1/3} \frac{k_{\rm b}}{\gamma - 1} \frac{\text{d}M_{\rm vir}}{\text{d}t},
\end{equation}
where $\gamma$ is the adiabatic index, and $\alpha$ is a coefficient that relates the virial mass and virial temperature, i.e. $T_{\rm vir} = \alpha M_{\rm vir}^{2/3}$ (see Equation \ref{eqn:tvir}).

\subsubsection{Atomic cooling limit}

Pop III star formation continues in DM halos until they are chemically enriched, prompting a transition to metal-enriched star formation that have more typical stellar masses.  Once the halo can cool through excited atomic hydrogen, \hh{} formation and therefore star formation \new{is much harder to suppress} by LW radiation because of the strong associated cooling rates \cite{Omukai01,Wolcott17}.  This is known as the atomic cooling limit and occurs at a virial temperature $T_{\rm vir} \simeq 10^4 \unit{K}$ when the cooling rates steeply rise by several orders of magnitude\cite{Sutherland93}.  The atomic cooling limit can be approximated with the function
\begin{equation}
    \label{eqn:acl}
    M_{\rm ACL} = 2 \times 10^7 \left( \frac{T_{\rm vir}}{10^4 \unit{K}} \right)^{1.5} \left( \frac{1+z}{21} \right)^{-1.5} \Ms
\end{equation}
that is calibrated with simulations.\cite{Fernandez14}

\subsection{Gravitational collapse and fragmentation}

Molecular hydrogen formation is relatively slow compared to chemistry in a metal-enriched dusty gas.  Furthermore, the \hh{} transitions are only energetic enough to cool to $\sim$200~K, compared to 10~K temperatures found in Galactic molecular clouds.  HD cooling can cool the gas down to $\sim$50~K.  The temperature plays an important role in determining the nature of star formation through the Jeans mass
\begin{equation}
    \label{eqn:jeans}
    M_{\rm J} = \frac{\pi^{5/2}}{6} \frac{c_s^3}{(G^3 \rho)^{1/2}}
\end{equation}
that describes the required mass of a gas cloud at a temperature $T$ that will gravitationally collapse.  Here $\rho$ and $c_{\rm s} = (\gamma k_{\rm b} T / \mu m_{\rm H})^{1/2}$ are the gas density and sound speed, respectively.  $\gamma$ is the adiabatic index.  Therefore given a metal-free and metal-enriched gas cloud at the same density, the metal-free one will have a Jeans mass a factor of $(200\unit{K}/10\unit{K})^{3/2} \simeq 90$ larger.  

To first order, this inefficient cooling of metal-free gas during collapse limits Jeans-mass fragmentation to clumps of no less than 1000 \Ms{}.  Within these large clumps, simulations have shown that only one or two dense cores will form\cite{Turk09} and then be able accrete from this reservoir with no or little competition.  As these cores continue to collapse, a central rapidly growing protostar forms.  It is surrounded by a protostellar accretion disk that can become gravitationally unstable, i.e. a Toomre parameter
\begin{equation}
    Q = \frac{c_s \kappa}{\pi G \Sigma} < 1.
\end{equation}
Here $\kappa$ is the epicyclic frequency, and $\Sigma$ is the gas surface density.  Recent simulations have between 20--50\% of the fragments are ejected from the disk before they can merge with the central protostar\cite{Greif12,Hirano17_P3,Susa19,Prole22}.

\subsection{Initial mass function and stellar endpoints}
\label{sec:ch9:imf}

The stellar initial mass function (IMF) is seemingly universal in which low-mass stars vastly outnumber high-mass stars\cite{Kroupa02}.  However by limiting gas cooling in metal-free gas, Pop III star formation stands as a potential exception where massive stars are predominant.  Obtaining stellar masses from cosmological simulations of Pop III star formation is difficult because of the needed high dynamic range and treatment of radiation transport.  The latter aspect arises from the fact that protostellar radiation is the process that ends the formation process.  Nevertheless, the bulk of the simulations in the past two decades have favored massive ($\gsim 8 \Ms$) star formation being the norm.  Low-mass star formation may still occur through the fragmentation of the protostellar disk as discussed in the previous subsection.

Although theoretical and observational campaigns have not well constrained the shape of the primordial stellar IMF, it is generally accepted that it is top-heavy, providing a likely source of SMBH seeds.  The Pop III IMF plays a very important role in the transition from primordial star formation to galaxy formation because it determines how often each stellar endpoint (e.g. BH, neutron stars, supernova) occurs.  Table \ref{tab:endpoints} lists the stellar mass ranges that have common endpoints and, if any, supernova types.  \new{For the interested reader, Chapter 3 details the astrophysical processes of stellar endpoints.}

\begin{table}[b]
    \tbl{Population III stellar endpoints \new{from non-rotating models\cite{Heger03}}}
    {\begin{tabular}[h]{@{}cccc@{}} \toprule
        \textbf{Approximate} & & & \\
        \textbf{mass range} & \textbf{Remnant} & \textbf{SN type} & \textbf{SN energy}\\
        (M$_\odot$) & & & (erg)\\
        \hline
        $< 9$ & White dwarf & None & --\\
        9--25 & Neutron star & Type II & $10^{51}$\\
        25--100 & Black hole & Weak Type II or None$^{\rm a}$ & $10^{49} - 10^{51}$\\
        100--140 & Black hole & Pulsational pair & $\sim 10^{51}$ per pulse\\
        140--260 & None & Pair-instability & $10^{51} - 10^{53}$\\
        $> 260$ & Black hole & None & --\\
        \botrule
    \end{tabular}}
    \begin{tabnote}
        $^{\rm a}$ Under some conditions, the star will directly collapse into a black hole without a supernova.
    \end{tabnote}
    \label{tab:endpoints}
\end{table}

Given an IMF $f_{\rm IMF}$ in terms of stellar mass per logarithmic mass, we can determine the fraction $f_{\rm rem}$ of stars with compact remnants, such as neutron stars and black holes with the following equation
\begin{equation}
    f_{\rm rem} = \frac{\int_{M_{\rm lo}}^{M_{\rm up}} f_{\rm IMF} \; \frac{dM}{M}}{\int_{0}^{\infty} f_{\rm IMF}\; \frac{dM}{M}},
\end{equation}
where $M_{\rm lo}$ and $M_{\rm up}$ are the lower and upper mass limits of a particular stellar endpoint.  For example, we can take the IMF to be a power-law with a slope of --1.3 that is near to the high-mass slope of the universal IMF.  We modify it with an exponential cut-off below a characteristic mass $M_{\rm char} = 20 \Ms$ with the function form
\begin{equation}
    f(\log M) dM = M^{-1.3} \exp\left[ -\left( \frac{M_{\rm char}}{M} \right)^{1.6} \right] \; dM,
\end{equation}
and $[M_{\rm lo}, M_{\rm up}] = [25, 140] \Ms$ for Pop III stars that leave a BH remnant\cite{Heger03}, 52 percent of Pop III stars will leave such a remnant.

\subsection{Supermassive stars}
\label{sec:ch9:sms}

Supermassive ($M > 1000 \Ms$) stars have long been theorized since the 1960's, originally as energy source of the then newly discovered quasars\cite{Fowler64}.  In the past two decades, there has been a renewed interest in supermassive stars as SMBH seeds.  Stellar evolution models and cosmological simulations have shown that a star can exceed $1000 \Ms$ if it is accreting at $\dot{M}_{\rm SMS} \gsim 0.04 \Ms$ per year\cite{Hosokawa11} that is a factor of 100--1000 greater than ones found in typically massive star formation.\cite{Krumholz14}

Two barriers to supermassive star formation exist in primordial star formation: low-mass fragmentation and high accretion rates.
\begin{description}
    \item[Fragmentation] would result in a stellar cluster with typically massive stars, effectively preventing supermassive protostellar cores from ever forming.  To prevent fragmentation below the supermassive scale, the gas must not cool below $\sim 10^4 \unit{K}$.  This lower limit suggests that the gas cannot cool through \hh{} or metal-line cooling, requiring \textit{strong \hh{} dissociation in a metal-free gas}.
    \item[High accretion rates] can only be supported in a strongly cooling gas in a deep gravitational potential well.\cite{Loeb94}  These conditions are realized in \textit{atomic cooling halos} (Equation \ref{eqn:acl}).  Protostars accreting above $\dot{M}_{\rm SMS}$ never come into equilibrium and never migrate to the main sequence.  They thus remain on the protostellar Hayashi track in the HR diagram with bloated envelopes and surface temperatures around 5000~K.  These temperatures are too low to produce sufficient ionizing radiation that usually halt accretion onto massive protostars.\cite{Hosokawa11}.
\end{description}

The most likely sites of supermassive star formation, if at all, are metal-free halos above the atomic cooling limit.  These halos must have had their star formation completely suppressed \new{and not be chemically enriched by supernovae} until it reaches $M_{\rm ACL}$.  Three plausible scenarios to suppress star formation are 
\begin{enumerate}
    \item being highly irradiated by LW radiation from a nearby galaxy, known as the close pair scenario, preventing \hh{} cooling,\cite{Dijkstra08, Visbal14, Regan16}
    \item dynamical heating (see Section \ref{sec:halos}) in a rapidly growing halo prevents \hh{} cooling and equilibration as it undergoes a set of major mergers\cite{Fernandez14,Wise19}, or
    \item the supersonic streaming velocities\cite{Tselia10} that exist between DM and baryons from the early universe are greater than the circular velocity $V_{\rm c}$ of the halo, preventing gas from condensing in the halo center.\cite{Hirano17}
\end{enumerate}

\section{Supermassive black hole seeding and early growth}
\label{ch9:sec:seeds}

% 10 pages
% In each subsection, I’ll discuss their formation conditions, cosmological environments, and number densities estimates
% a. Primordial black holes
% b. Light seeds (< 100 Msun): stellar remnants from primordial stars
% c. Intermediate mass seeds (100 – 10,000 Msun): remnants of very massive stars, formation in nuclear star clusters
% d. Massive seeds (>10,000 Msun): remnants of supermassive stars

\begin{figure}[t]
    \centering
    \includegraphics[width=\textwidth]{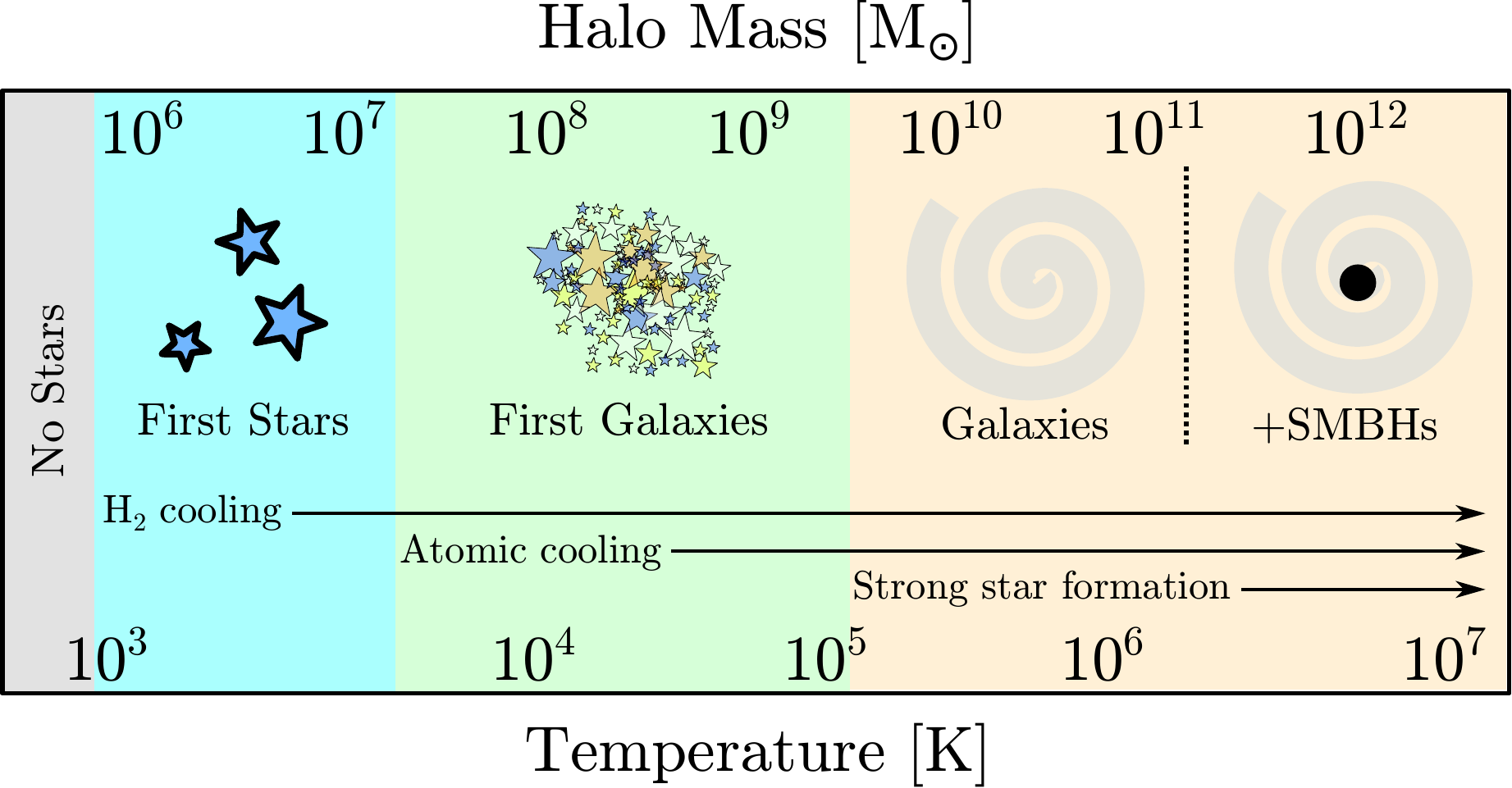}
    \caption{\label{fig:landscape}\textbf{Landscape of early galaxy and black hole formation.}  As halos grow, they host increasingly more cold gas and fuel strong star formation, ranging from primordial (Pop III) stars ($10^{5.5} - 10^{7.5} \Ms$) to the first generations of galaxies ($10^{7.5} - 10^{9.5} \Ms$) to more massive galaxies and SMBHs.  Massive BH seeds form in the less massive atomic cooling halos.}
\end{figure}

The initial growth of the first BHs is highly dependent on its stellar progenitor and how it affected the adjacent environment that will later accrete onto the BH seed.  Afterward cosmological effects, such as mergers, play a significant role in the evolution of the seed BH and its host galaxies.  As the galaxy grows, both star formation and SMBH fueling will become self-regulated through their feedback processes.\cite{Milo09_IMBH, Park11_IMBH1}

There are three favored seeding scenarios that all originate from Pop III stellar remnants (see Section \ref{ch9:sec:pop3}) but differ in their progenitor masses.
\begin{enumerate}
    \item A typically massive star (Section \ref{sec:ch9:imf}) with a mass on the order of 10 \Ms{}.
    \item A very massive star with a mass on the order of 1000 \Ms{}, resulting from stellar collisions in a dense stellar cluster.\cite{Begelman78}
    \item A supermassive star (Section \ref{sec:ch9:sms}) with a mass $\gsim 10^4 \Ms$, forming from its direct collapse.
\end{enumerate}
As discussed previously, the last two scenarios are restricted to host halos above the atomic cooling limit (Equation \ref{eqn:acl}).  This critical halo mass can be thought as the transition point between isolated Pop III star formation and the first generations of galaxies, illustrated in Figure \ref{fig:landscape}.

\subsection{Light seeds}

\begin{figure}[t]
    \centering
    \includegraphics[width=\textwidth]{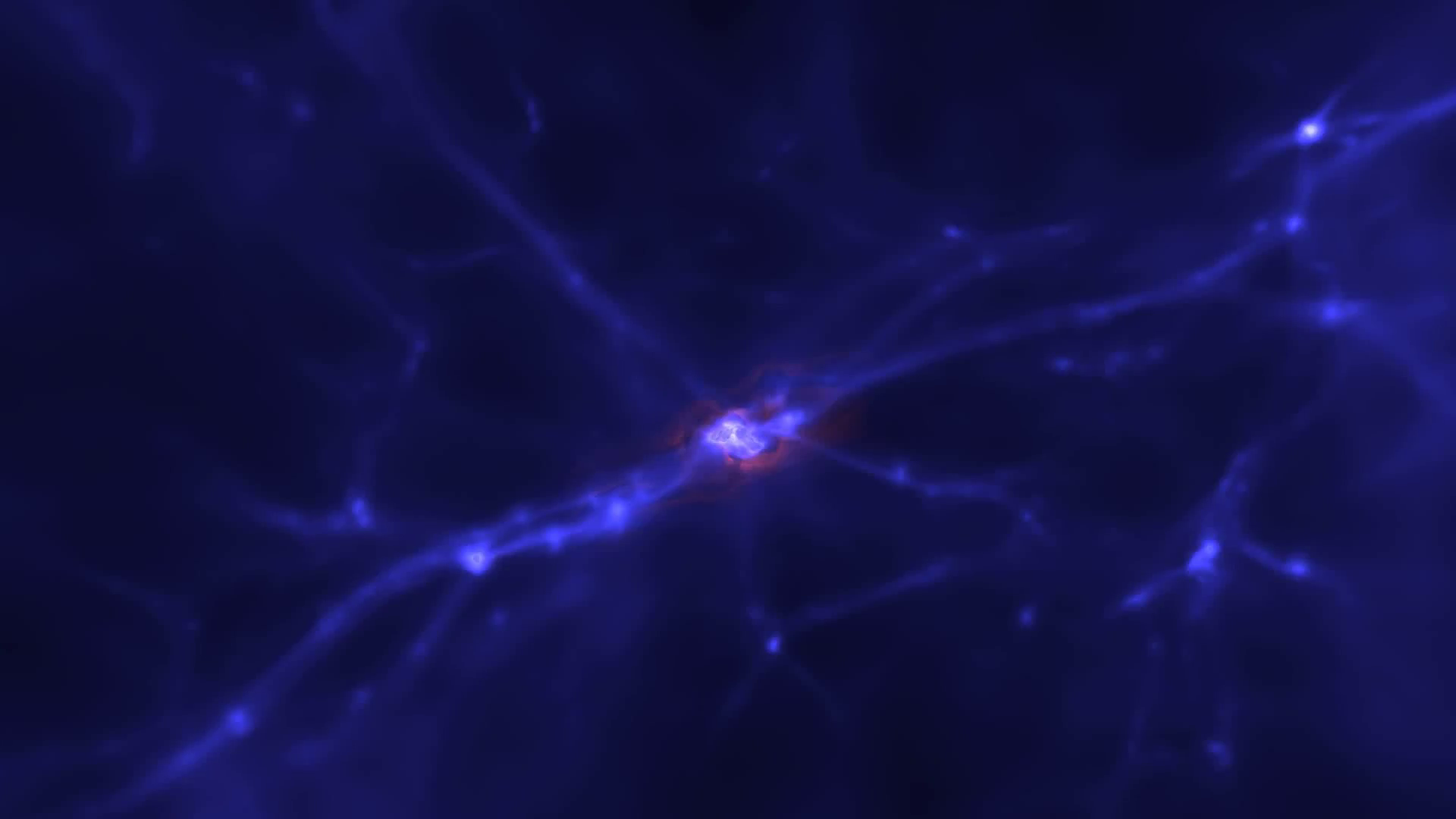}
    \includegraphics[width=\textwidth]{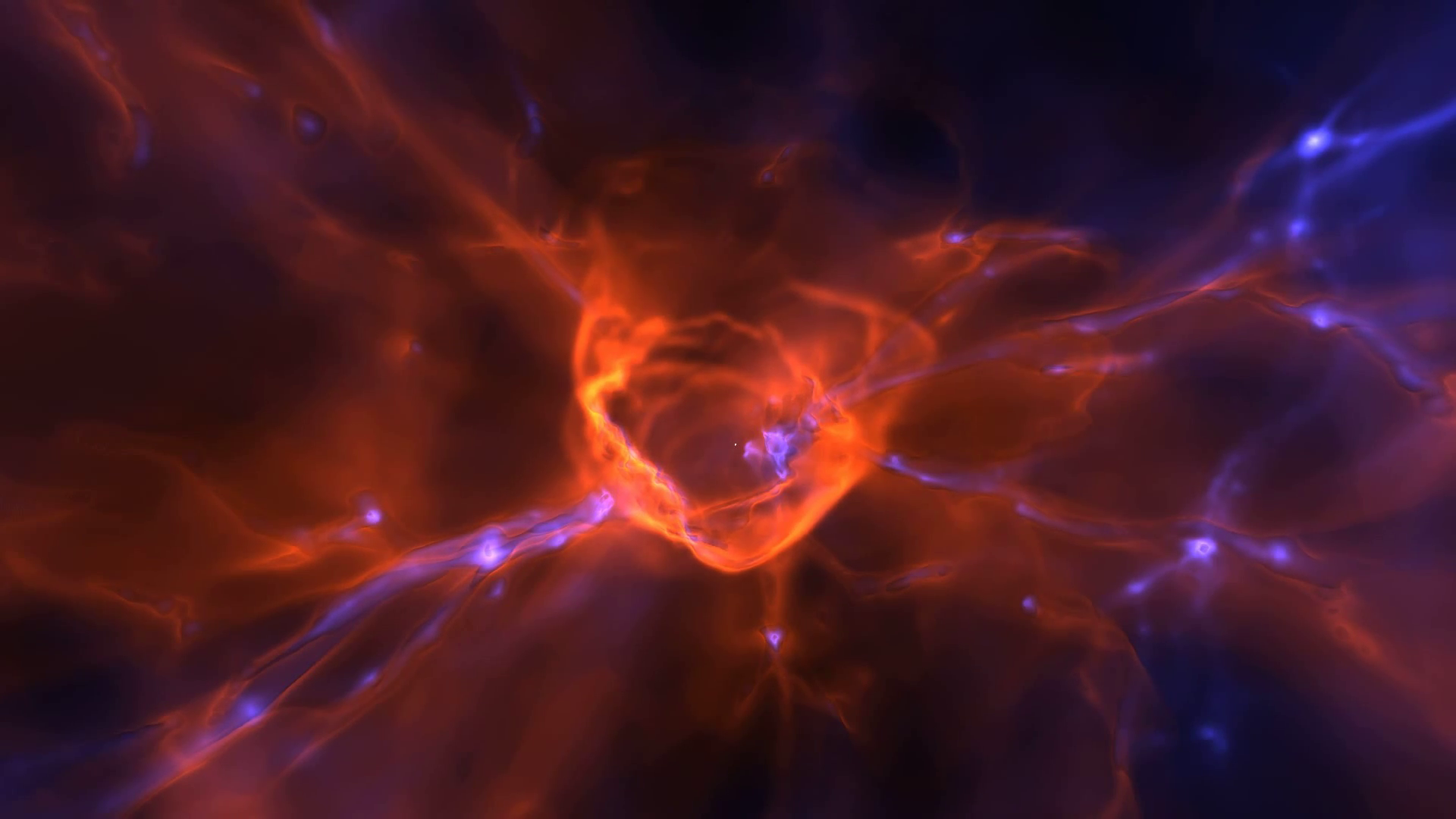}
    \caption{\label{fig:p3}\textbf{Simulated regions around a Pop III star.} These volume renderings of a cosmological simulation of Pop III star formation depict the cosmological structure before star formation (\textit{top}) and after its supernova (\textit{bottom}).  The field of view is approximately 2 kpc and is centered on the star and host halo at $z \simeq 20$.  The blue and orange colors represent cold (100~K) and warm ($10^4 \unit{K}$) gas, respectively.  The image brightness is scaled with gas density. {\it Credit:} R. K{\"a}hler, J. Wise, \& T. Abel.}
\end{figure}

Pop III stars form mostly in $\sim 10^6 \Ms$ minihalos.  During their main sequence, most of the gas in the host halo is swept up by the $\sim 30 \kms$ shock associated with the ionization front.  This creates a nebula that contains a warm ($10^4\unit{K}$) and diffuse (1~\cubecm) medium\cite{kitayama04,Whalen04,Abel07}.  Figure \ref{fig:p3} shows a volume rendering of a simulation with Pop III star formation, radiative feedback during its main sequence, and the subsequent supernova.  In the top panel, the host halo is unaffected before star formation.  In the bottom panel, the host halo is completely disrupted by the ultraviolet radiation heating and supernova explosion.  This environment is not conducive for BH growth, and it must wait until a gas-rich merger to resupply the gas reservoir in the host halo.  During which time, the seed does not \new{grow} more than 1 percent.\cite{Alvarez09}

Simulations following the growth of these light seed BHs have shown that a cold gas reservoir is not re-established until the halo reaches the atomic cooling limit.  Only then, the BH can accrete at a non-negligible rate.\cite{Jeon12}  If the BH remains in the galaxy outskirts, it does not encounter much dense gas to accrete because it is continuously depleted through star formation.\cite{Smith18}

\subsection{Intermediate mass seeds}

Stellar collisions and dynamical instabilities can cause dense stellar clusters to form seed BHs within intermediate masses\cite{Omukai08, Devecchi09, Katz15, Tagawa20, Schleicher22}.  Regardless of the formation mechanism, they all have massive stellar precursors that will photo-heat the immediate environment.  During this phase, dynamical effects between massive stars and/or BHs ($\sim 10 \Ms$) will drive dense gas inflows toward the center.  This inflow fuels the seed BHs at very high rates.\cite{Davies11,Alexander14}

Afterward, further gas supply may be disrupted by stellar and BH feedback.  If they are hosted in a minihalo, they might have a difficult time \new{growing} without external gas supplies, similar to the light seeds.  However being embedded in a stellar cluster with $M_\star \simeq 10^4 \Ms$\cite{Katz15,Sakurai19}, the deeper potential well could mitigate any fueling issues.\cite{Park16}  If the host halo is above the atomic cooling limit (Equation \ref{eqn:acl}), the seed BH should avoid these initial growing pains if located in the halo center.  Most of the research to-date have focused on the initial formation and not so much on its early subsequent growth.

\subsection{Heavy seeds}

Supermassive stars with masses over $10^4 \Ms$ will undergo a gravitational, sometimes relativistic, instability.  This leads to the creation of a massive BH with nearly all of the stellar mass trapped in it.  These hypothetical objects can grow up to a mass of $10^6 \Ms$ with $\sim 10$ percent collapsing into a BH known as a direct collapse black hole (DCBH)\cite{Begelman06,Begelman08,Johnson13_DCBH}.

Direct collapses into BHs occur for any initial stellar mass above $\sim 260 \Ms$\cite{Heger03,Begelman06} with the exception of a narrow range around 55{,}000 \Ms{}.  This rare hypothetical case produces a tremendous supernova explosion at nearly $10^{55} \unit{erg}$ that is $10^4$ times more energetic than a normal core-collapse supernova.\cite{Montero12, Whalen13_SMS}.  During the collapse or explosion process, some fraction of supermassive stars can launch a relativistic jet that powers a gamma-ray burst.\cite{Matsumoto15}

These supermassive stars form in the centers of atomic cooling halos and are surrounded by a large gas supply.  It is not disrupted, like in the light and intermediate seed cases, from stellar radiation from its progenitor because of its low surface temperature ($\sim 5000 \unit{K}$) and thus low ionizing luminosity.  However after the BH remnant is created, it should accrete rapidly and emit strongly in the ultraviolet and X-rays.  Cosmological simulations have shown that this radiation may trigger star formation within tens of parsecs.  This would create a luminous object with an overly massive BH relative to the surrounding galaxy, unique to the high-redshift universe.\cite{Agarwal13,Barrow18}

A hybrid scenario with a mixed population of BH seeds within a single halo is also plausible.  As the halo gravitationally collapses, \hh{} will form in the center, becoming self-shielding and cooling to hundreds of Kelvin.  Recent simulations have shown such a cluster of both typically massive Pop III stars and supermassive stars could form in a metal-free atomic cooling halo.\cite{Regan20_VMS}  The cluster has a radius of $\sim 10\unit{pc}$, and a fraction of the BH remnants could merge at a later time, creating gravitational waves in the process.

\subsection{Primordial black holes}

BHs could also form in the very early universe during the first second after the Big Bang.\cite{Carr21}  These BHs are unlike the other seeds because they are not connected with any stellar endpoints.  Large-scale structure is thought to be seeded by quantum fluctuations in the inflationary era that ended around $10^{-32}$ seconds after the Big Bang.  These fluctuations are well modeled by a Gaussian random field, thus there exists a non-zero probability that some regions will have a density $\rho$ high enough to create BH, specifically the escape velocity needs to be superluminal
\begin{equation}
    v_{\rm esc} = \sqrt{\frac{2GM}{r}} = \sqrt{\frac{8\pi G \rho r^2}{3}} > c,
\end{equation}
where $M$ is the mass enclosed in a sphere of radius $r$.  Because the density \new{during the radiation-dominated era} is decreasing with time ($\rho \propto t^{-2}$) as the universe expands, a primordial BH would have a mass on the order
\begin{equation}
    M_{\rm BH} \sim \frac{c^3 t}{G} \sim 10^{15} \left( \frac{t}{10^{-23} \unit{s}} \right) \unit{g}
\end{equation}
at a time $t$ after the Big Bang.  Primordial BHs could theoretically have a wide mass distribution, spanning from the Planck mass ($10^{-5} \unit{g}$) if formed at the Planck time ($10^{-43} \unit{s}$) to ones rivaling SMBHs with $10^5 \Ms$ if formed at $t = 1 \unit{s}$.

In addition to forming through inhomogeneities, primordial BHs may form during cosmic phase transitions.  These are associated with the separation of the fundamental forces as the mass-energy density of the universe decreases, for example, the electroweak force separating into the electromagnetic and weak forces at $t = 10^{-12} \unit{s}$.  Any symmetry breaking during these transitions could produce space-time defects from which primordial BHs form.\cite{Carr21}.

There are both theoretical and observational constraints on their masses.  First, all BHs emit a small amount of radiation, known as Hawking radiation, and thus slowly lose mass over time and eventually evaporate.  The BH evaporation timescale is proportional to its mass cubed.  This timescale is equal to the current age of the Universe for a $10^{15} \unit{g}$ BH.  Therefore, any primordial BH below this mass scale would have already evaporated.  Second, there are several observational constraints on their masses: the lack of gravitational microlensing events, which happen when a non-luminous massive object passes between the observer and the light source; dynamical effects from their interactions with stars in globular clusters and galaxies; large-scale structure and cosmological background constraints\cite{Poulin17}.  These constraints limit their possible masses to \new{three} ranges: asteroid masses ($10^{16} - 10^{17} \unit{g}$), sublunar masses ($10^{20} - 10^{26} \unit{g}$), \new{and massive primordial BHs ($\sim 10^3 - 10^5 \Ms$)}.  Because these allowed mass ranges are significantly smaller than the stellar remnant BH masses, they are usually not considered as likely seeds of SMBHs.

\section{Host dark matter halos and timescale estimates}
\label{ch9:halos}

We begin by exploring BH growth by connecting the abundance of SMBHs at $z \sim 6$ with the abundance of DM halos, known as abundance matching.  These BHs already have masses up to $10^9 \Ms$ when the universe is only 900 Myr.  By knowing their DM host halo mass, we can calculate their typical growth history with a specific focus on when they could form Pop III stars and seed BHs.  This gives a timescale in which the BHs can grow from their seed masses to their observed masses at $z \sim 6$.  We first need to determine the halo mass function that provides the abundances of DM halos.

\subsection{Dark matter halo abundances}

The random Gaussian nature of cosmological density fluctuations allows us to calculate the abundance of DM halos at some mass $M$ and redshift $z$, known as Press-Schechter (PS) formalism.  The general idea is to combine a model of spherical collapse within a cosmological context with the linear growth of large-scale structure.  In the spherical collapse model, objects collapse and form a virialized DM halo when the linearly extrapolated overdensity reaches $\delta > \delta_{\rm c} \simeq 1.69$.  To consider different halo masses, the overdensity field is smoothed with a window function $W$ (e.g. top-hat, Gaussian, sharp-k) with radius $R$ that corresponds to a mass $M = \gamma_{\rm f} \rho R^3$, where $\gamma_{\rm f}$ depends on the window function and is of order unity and $\rho$ is the mean matter density of the universe at redshift $z$.

\begin{figure}[t]
    \centering
    \includegraphics[width=0.5\textwidth]{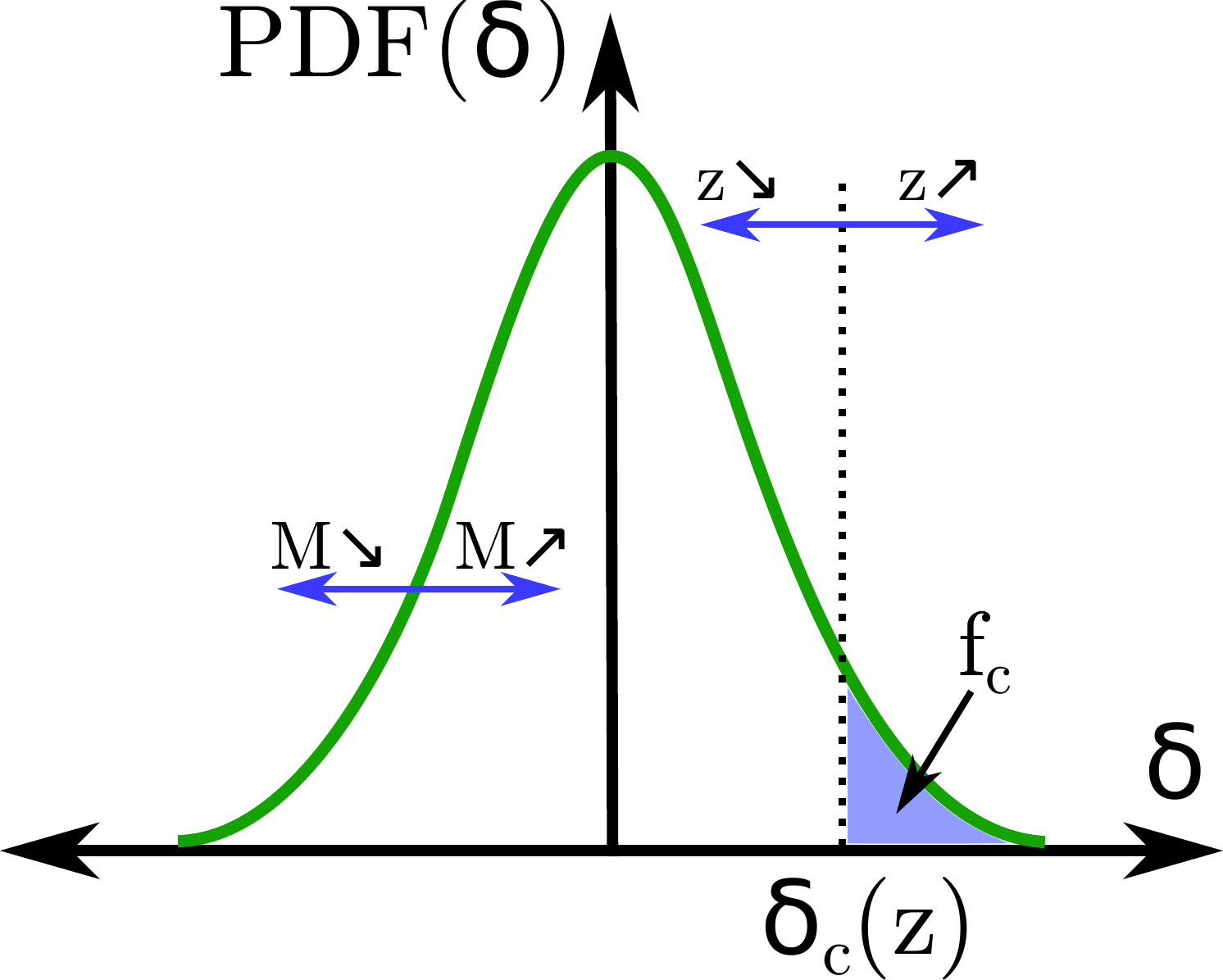}
    \caption{\label{fig:halos}\textbf{Probability distribution function of density fluctuations.}  Dark matter halo abundances can be computed by combining linear perturbation theory of structure formation and spherical collapse models.  A mass fraction $f_{\rm c}$ with a linearly extrapolated density above $\delta_{\rm c}$ is assumed to be contained in a DM halo. The distribution becomes narrower with increasing halo mass $M$ and the critical overdensity $\delta_{\rm c}$ decreases with redshift.}
\end{figure}

The ansatz of PS formalism is that the probability that smoothed overdensity $\delta_{\rm s} > \delta_{\rm c}(t)$ is the same as the mass fraction contained in halos with a mass greater than $M$, which is illustrated in Figure \ref{fig:halos}.  Thus the probability that some mass exists in a viralized halo with mass $M$ is given by the mass fraction above $\delta_{\rm c}$ at time $t$,
\begin{equation}
    \mathcal{P}[> \delta_{\rm c}(t)] = \frac{1}{\sqrt{2\pi} \sigma(M)} \int_{\delta_{\rm c}(t)}^{\infty} \exp \left[ - \frac{\delta_{\rm s}^2}{2\sigma^2(M)} \right] d\delta_{\rm s} = \frac{1}{2} \text{erfc} \left[ \frac{\delta_{\rm c}(t)}{\sqrt{2} \sigma(M)} \right],
\end{equation}
where
\begin{equation}
    \sigma^2(M) = \langle \delta_{\rm s}^2(\mathbf{x}; R) \rangle = \frac{1}{2\pi^2} \int_0^\infty P(k) \widetilde{W}^2(kR) k^2 \text{d}k
\end{equation}
is the mass variance of the smoothed density field with $P(k)$ as the power spectrum of the density perturbations.  It is normalized to $\sigma_8$ at a radius of $8 h^{-1} \unit{Mpc}$, usually measured through the cosmic microwave background and galaxy clusters. Here $\mathbf{x}$ is the location; $k$ is the wavenumber; $\widetilde{W}$ is the Fourier transform of the window function.

PS formalism suffers from the ``cloud-in-cloud'' problem that occurs when an underdense region does not collapse in isolation.  However if it is contained within some larger collapsing region, it could collapse.  Therefore, the collapse mass fraction is given by
\begin{equation}
    F(>M) = 2\mathcal{P}[> \delta_{\rm c}(t)].
\end{equation}
\new{See Section 5.8 of ref. \citeauthor{PadmanabhanBook} for an insightful discussion on the factor of two.}  We use this mass fraction to calculate the number density $n$ of DM halos between some mass range $(M, M+dM)$ by differentiating it with respect to halo mass
\begin{equation}
    \begin{aligned}
    n(M,t) \text{d}M &= \frac{\rho}{M} \frac{\partial F(>M)}{\partial M} \text{d}M\\
    &= \sqrt{\frac{2}{\pi}} \frac{\rho}{M^2} \frac{\delta_{\rm c}}{\sigma} \exp\left( -\frac{\delta_{\rm c}^2}{2\sigma^2} \right) \left| \frac{\text{d}\ln \sigma}{\text{d}\ln M} \right| \text{d}M.
    \end{aligned}
\end{equation}
For convenience, we define the variable $\nu = \delta_{\rm c}(t)/\sigma(M)$ that signifies the halo rareness and its mass within Gaussian statistics.  For example, a halo with $\nu = 2$ would be referred to as a 2-$\sigma$ halo or peak, and given a time $t$ and cosmology, the mass can be computed from $\sigma(M)$.

\subsection{Abundance matching and growth timescales}

\begin{figure}[t]
    \centering
    \includegraphics[width=0.48\textwidth]{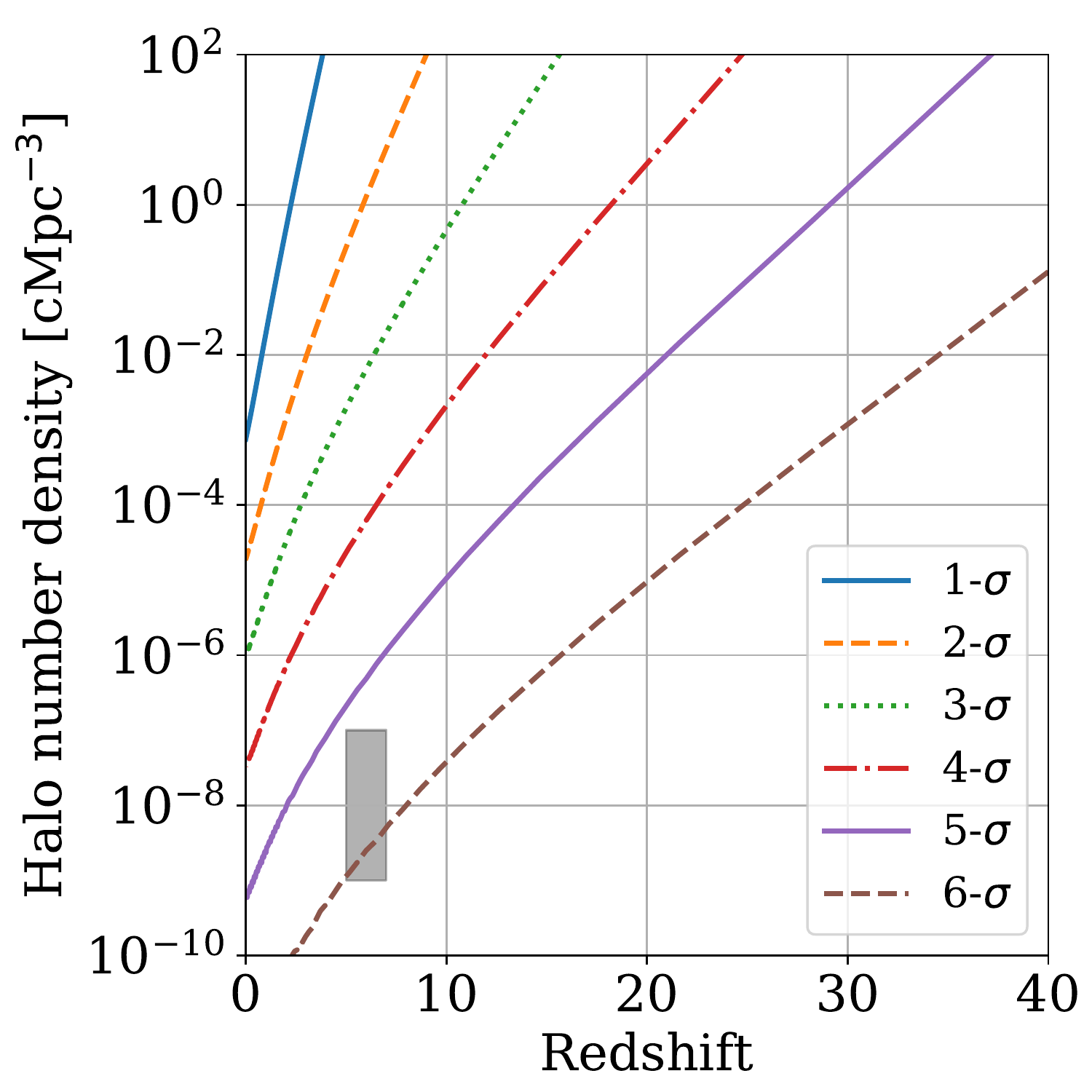}
    \hfill
    \includegraphics[width=0.48\textwidth]{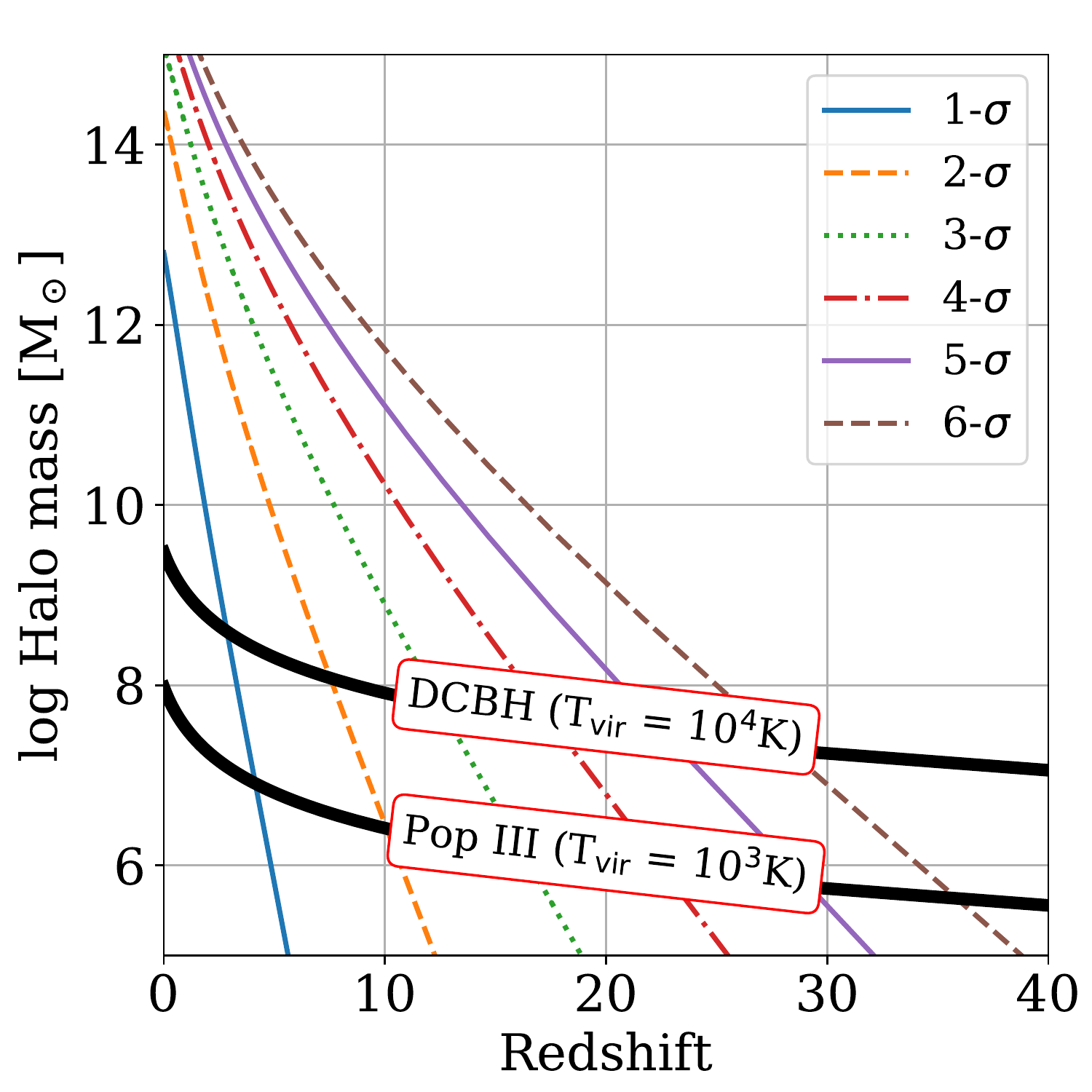}
    \caption{\label{fig:abundance}\textbf{Abundance matching of supermassive black holes at redshift 6.} \textit{Left:} Each line shows the comoving number density of DM halos for a particular rarity $\nu$.  The shaded box depicts an estimate of the SMBH number density, considering a duty cycle between 1\% and 100\%.  Simple abundance matching suggests that their host halos are very rare halos ($5-6 \sigma$).  \textit{Right:} DM halos masses for particular rarities compared with critical halo masses, obtained from their virial temperatures (Equation \ref{eqn:tvir}), for Pop III star formation and direct collapse BH (DCBH) formation from supermassive stars.  The 5- and 6-$\sigma$ tracks can be used to estimate the progenitor halo masses of the $z \sim 6$ hosts at earlier redshifts.}
\end{figure}

The observed abundance of $z \sim 6$ active galactic nuclei (AGN) is approximately one per comoving Gpc.  However, only some fraction of SMBH are in the active state that is around 10 percent.\cite{Luo11}  Giving an order of magnitude leeway in the SMBH abundance, we can compare the SMBH comoving number density of $10^{-7} - 10^{-9}$ per comoving Mpc$^3$ to the halo number density of various halos with $\nu$, shown in Figure \ref{fig:abundance}.

The left panel shows the number density of DM halos with a rarity $\nu$ from Gaussian statistics, as defined in the previous subsection.  The grey shaded \new{area} shows the estimated number density of SMBHs at $z \sim 6$.  By comparing these two abundances, we conclude that these most massive SMBHs are hosted by very rare halos with $\nu = 5-6$.  Now we can look at the associated halo masses of these $\nu$ values, as shown in the right panel of Figure \ref{fig:abundance}.  These halos have masses around $10^{13} \Ms$ at $z \sim 6$, ten times more massive than the Milky Way halo, showing that they have fiercely assembled within the early universe.  \new{However, ALMA observations that resolve galaxies hosting $z \sim 6$ quasars indicate that their dynamical masses are an order of magnitude below low-redshift relations \cite{Wang16, Decarli18, Shimasaku19}.  It should be noted that these interstellar medium tracers can underestimate dynamical masses, and these observed quasars may suffer from a selection bias, identifying only the brightest objects \cite{Volonteri11, Lupi19}.}

We can then trace back the 5-$\sigma$ and 6-$\sigma$ halo masses to higher redshifts to estimate a typical growth history of the most massive halo progenitor.  The right panel also shows the halo masses associated with the critical virial temperatures needed to form Pop III stars, which can provide light seeds, and heavy seeds, known as direct collapse black holes (DCBHs), that form from supermassive stars in atomic cooling halos.  The redshift at which these the halo masses cross these thresholds provide an estimate of their formation time.  We thus conclude that the $10^9 \Ms$ SMBH host halos at $z \sim 6$ must have formed its first Pop III stars around $z = 30-35$.  It then crossed the atomic cooling limit around $z = 25-30$.  At these redshifts, the universe is around 100 million years old.  The seed BHs thus only have 800 million years to grow by $4-7$ orders of magnitude so that they transform into the observed SMBHs at $z \sim 6$.  Aiding this difficult ascent, their host halos will grow by a similar factor.

\section{Black hole growth during the Epoch of Reionization}
\label{ch9:sec:grow}

% 10 pages
% a. Light seeds: inefficient growth in most cases but super-Eddington accretion is possible under specific conditions. Weak dynamical friction → most likely not in halo centers
% b. IMBH seeds: growth and interactions within dense clusters and incorporation into larger halos. (Less done in this topic than the light and massive seeds.)
% c. Massive seeds: possible rapid growth in gas-rich environments after formation from lack of SMS feedback. Some form in the very rare peaks with high halo growth and BH accretion rates. Metal-free star formation around massive seeds?

Any BH growing in a galaxy must compete with star formation and feedback processes for any accretion.  Through a sequence of mergers and smooth accretion from the intergalactic medium, the halo will have an ample gas supply once it crosses the atomic cooling limit.  The key difficulty for any BH accretion is the long journey from the halo virial radius to the BH event horizon, affected by many physical processes along the way.

\subsection{Black hole accretion}

Any gas that crosses the innermost stable circular orbit ($R_{\rm ISCO} = 6GM_{\rm BH}/c^2$) of the BH will be accreted.  However, it must first shed most of its angular momentum before sinking to such small radii.  The gas inside pre-galactic halos is mainly turbulent as it comes into virial equilibrium\cite{Wise07,Greif08} and has some coherent rotation with speeds just under half of the circular velocity $V_{\rm c} = \sqrt{GM/R}$.\cite{Regan09}  Being a turbulent medium, some fraction of gas will have a very low specific angular momentum $|\mathbf{j}| = |\mathbf{v} \times \mathbf{r}|$, either already existing at small radii or on nearly radial orbits.  This gas is able to migrate near (pc-scale) the BH without being rotationally supported.\cite{Lodato06,Lodato07}

\subsubsection{Bondi-Hoyle accretion}

Once the gas enters the central regions of the halo, the next barrier to accretion is becoming gravitationally bound to the BH.  This was first described in the context of a point mass in a spherically symmetric system by Bondi, Hoyle and Lyttleton in the 1940s and early 1950s.\cite{Bondi44, Bondi47, Bondi52}  They calculated the inflow rates of matter onto this point mass, where thermal pressure is the only force counterbalancing gravity.  The radius at which the gravitational force is greater than thermal pressure forces is called the Bondi radius $r_{\rm B}$.  A monoatomic gas with an adiabatic equation of state and an adiabatic index $\gamma = 5/3$ has a Bondi radius
\begin{equation}
    \begin{aligned}
    r_{\rm B} &\equiv \frac{GM_{\rm BH}}{c_{\rm s}^2}\\
    &= 0.31 \left( \frac{M_{\rm BH}}{100 \Ms} \right) \left( \frac{T}{100 \unit{K}} \right)^{-1} \unit{pc}.
    \end{aligned}
\end{equation}
One can make an analogy to the escape velocity, but here sound waves traveling at $c_{\rm s}$ are not fast enough inside the Bondi radius to equilibrate the system through pressure waves.  The gas cloud is free to collapse and accrete onto the BH.

Once the gas enters the Bondi radius, the point mass is assumed to accrete at the Bondi-Hoyle rate\footnote{{\it nb.} The ``BH'' subscript refers to black hole.}
\begin{equation}
    \label{eqn:bondi_rate}
    \begin{aligned}
    \dot{M}_{\rm BH} &= \frac{4\pi (G M_{\rm BH})^2 \rho}{c_{\rm s}^3}\\
    &= 2.18 \times 10^{-8} \left( \frac{M_{\rm BH}}{100 \Ms} \right)^2 \left( \frac{\rho}{10^{-24} \unit{g cm}^{-3}} \right) \left( \frac{T}{100 \unit{K}} \right)^{-3/2} \Ms \; \text{yr}^{-1}
    \end{aligned}
\end{equation}
where $\rho$ is the gas density at the Bondi radius.  If the BH is moving relative to the ambient gas, it is more difficult to gravitationally capture the gas.  This factors into an effective sound speed by adding the relative velocity $v_{\rm rel}$ in quadrature to the sound speed,
\begin{equation}
    \dot{M}_{\rm BH} = \frac{4\pi (G M_{\rm BH})^2 \rho}{(c_{\rm s}^2 + v_{\rm rel}^2)^{3/2}}.
\end{equation}

In reality, the BH exists in an external gravitational potential that could aid in accretion.  Sources of this potential are DM that has a cusped density profile and a stellar bulge.  These additional sources of gravitational potential effectively increases the Bondi radius.  The most significant increase occurs when a light seed BH ($\sim 100 \Ms$) resides in a stellar bulge with a mass exceeding $10^6 \Ms$, resulting in an increase of $10^4$ in the effective Bondi radius and a similar factor for the BH accretion rate.\cite{Park16}

\subsubsection{Eddington-limited accretion}

Once the gas enters the (effective) Bondi radius, some fraction will deposit onto the accretion disk orbiting the BH.  The gas loses further angular momentum through viscous forces and magnetic instabilities.\cite{Balbus91,Power11}.  Eventually the gas will reach the inner edge of the accretion disk, and a fraction will be accreted into the BH. 

At this stage, the radiation emitted from the medium around the BH can apply an outward force, limiting any inflows.  This accretion limit and associated luminosity are known as the Eddington limit and luminosity, respectively.  First, we assume that the accreted gas emits some fraction $\epsilon$ of its rest-mass into radiation.  In spherical symmetry, the accretion will be halted if the radiation pressure due to Thomson scattering exceeds the gravitational force.  When these two forces are equal in this simplified scenario, the BH is accreting at its maximum
\begin{equation}
    \begin{aligned}
    \dot{M}_{\rm Edd} &= \frac{4\pi G M_{\rm BH} m_{\rm p}}{\epsilon c \sigma_{\rm T}},\\
    &= 2.22 \times 10^{-6} \left( \frac{M_{\rm BH}}{100 \Ms} \right) \left( \frac{\epsilon}{0.1} \right)^{-1} \Ms \; \text{yr}^{-1},
    \end{aligned}
\end{equation}
where $\sigma_{\rm T}$ is the Thomson scattering cross-section.  The Eddington luminosity can be expressed as
\begin{equation}
    \begin{aligned}
        L_{\rm Edd} &= \epsilon \dot{M}_{\rm Edd} c^2\\
        &= 1.26 \times 10^{40} \left( \frac{M_{\rm BH}}{100 \Ms} \right) \unit{erg s}^{-1}.
    \end{aligned}
\end{equation}

Because its growth rate is proportional to its mass, the growth is exponential with an $e$-folding time of $45 (\epsilon/0.1) \unit{Myr}$.  Applying this rate to the most massive SMBHs at $z \sim 6$, only 800 million years exist from their approximate seeding time to $z \sim 6$, which is nearly 18 $e$-folding times.  Thus the seed BH can grow by a factor of $e^{(800 \unit{Myr} /45 \unit{Myr})} = 10^{7.7}$ if it grows at the Eddington limit, which can be far from the case, considering the radiation and supernovae from nearby stars and the feedback from the BH itself.

\subsubsection{Super-Eddington accretion}

On the other hand, the Eddington limit is derived in spherical symmetry.  Accretion rates above this limit, known as super-Eddington accretion or hyper-accretion, can occur when the following two assumptions break down.
\begin{enumerate}
    \item The surrounding medium is not spherically symmetric or is porous.  This is likely the case as BHs are surrounded by accretion disks\cite{Abramowicz88, Novak12, Jiang14} and/or (supersonic) turbulence\cite{VanBorm13}.
    \item The radiation is trapped with the gas flow.  In radiation trapping, it is advected with the fluid flow and does not impart all of its momentum to the absorbing or scattering medium.\cite{Begelman79, Inayoshi16, Pacucci17}
\end{enumerate}
When any of these conditions occur, the BH accretion rates can easily exceed the Eddington rate, sometimes up to a factor of $\sim 200$ for short periods at a radiative efficiency $\epsilon \simeq 0.05$\cite{Jiang14}.  These hyper-accretion bursts partially alleviate the constraints on \new{the} growth history of the most massive SMBHs at high redshifts\cite{volonteri09}.

\subsection{Black hole feedback}

When a BH accretes gas, a fraction of its mass-energy will be returned to the surrounding environment.  What fraction is unclear, but given the high growth rates required to transform into the observed SMBHs, any fraction may have a substantial effect on future growth.  These feedback processes can be broadly categorized into two phases---radiative (quasar) and kinetic---that are determined by the accretion rates.  If the SMBH is weakly accreting at $\dot{M}_{\rm BH} \lsim 0.01 \dot{M}_{\rm Edd}$, it will not produce much radiation but will launch a jet.  Otherwise it will strongly radiate in the quasar phase.  Both types of feedback can launch winds from SMBH that are ubiquitous in AGN, measured from their absorption spectra.
\begin{description}
    \item[Radiative feedback] includes various radiative processes, such as blackbody emission and inverse Compton scattering in the disk and corona, respectively.  When the medium is optically thin, i.e. the photons stream through it, the luminosity scales as $L = \epsilon \dot{M}_{\rm BH} c^2$.  When the medium is optically thick, the photons scatter within the medium and is trapped, weakening the dependence on accretion rate, $L \propto \ln(\dot{M}_{\rm BH})$.  The resulting radiation can have two major impacts on the medium.
    \begin{itemize}
        \item {\it Photo-heating:} Any photons that ionize atoms will heat the gas from the excess photon energy $\Delta E = E_\gamma - E_i$ that goes into the free electron, where $E_i$ is the atomic ionization potential.  Its kinetic energy will quickly thermalize and heat the gas.  The increased temperature will reduce the Bondi-Hoyle accretion rate (Equation \ref{eqn:bondi_rate}).
        \item {\it Radiation pressure:} Any photons absorbed by the medium will impart their momenta $|\mathbf{p}| = E_\gamma/c$ to the absorbing species.  This force will decelerate or, in some cases, reverse the gas inflow toward the BH.
    \end{itemize}
    \item[Kinetic (jet) feedback] occurs when bipolar relativistic jets are launched near the BH horizon, releasing a significant amount of energy in the process.  The jets entrain some fraction of mass as it travels through interstellar space and eventually exiting the host galaxy.  If we assume that some fraction $\epsilon_{\rm kin}$ of the luminosity $L_{\rm BH}$ is coupled to the jet, we can write
        \begin{equation}
            L_{\rm BH} = P_{\rm kin} + (\text{rate of change in gravitational potential energy}).
        \end{equation}
        At any radius, some energy will be spent climbing out of the potential well.  We can encode this energy into the coupling constant $\epsilon_{\rm kin}$.  Now the jet ``kinetic'' power $P_{\rm kin}$ can be expressed as
        \begin{equation}
            P_{\rm kin} = \epsilon_{\rm kin} L_{\rm BH} = \frac{1}{2} \dot{M}_{\rm jet} v_{\rm jet}^2.
        \end{equation}
        The relativistic jet heats the nearby gas as it shocks the intervening medium.  From conservation of energy, it will slow to non-relativistic speeds yet still supersonic.  But eventually the outer jet will stall out at large radii.  The most powerful jets can travel upwards of a Mpc from its SMBH, creating large radio-emitting lobes.
\end{description}

\subsection{Incorporation into larger galaxies}

We have discussed that the observed $z \sim 6$ quasars are located in rare halos of mass $10^{13} \Ms$ that have rapid growth histories, forming their first stars at $z \sim 30$ and growing by a factor of a million within 800 million years.  These halos will be bombarded with infalling halos and gas accretion, fueling intense star formation and central BH growth.  In the process, they will have several major mergers and hundreds, if not thousands, of minor mergers with halos above the atomic cooling limit.  Each halo contains a first-generation galaxy and most likely a population of seed BHs.  More typical galaxies, like the Milky Way, grow at a more leisurely rate, whose progenitor halos can only support high star formation rates after $z \sim 10$.  If the central BH growth is stunted in these normal galaxies, it is not a fueling problem because they have billions of years to go to the AGN population observed at lower redshifts.  Nevertheless, theories of BH formation and growth should capture both the typical and extreme cases.  \new{For the interested reader, Chapter 5 discusses the following and related topics in detail.}

\subsubsection{Central black hole occupation fractions}

During the initial assembly of the galaxies just after reaching the atomic cooling limit, the galaxy is not guaranteed to have a central BH.  There is some probability that its Pop III stars do not leave a BH, such as ones that leave a neutron star or leave no remnant in a pair-instability supernova.  When the most massive progenitor does not have a central BH, the early galaxy will initially have BHs orbiting the galaxy outside the nuclear ($\sim 10 \unit{pc}$) region as their host halos merge into the primary galaxy.  Simulations have found that only 20 percent of high-redshift galaxies with stellar masses around $10^6 \Ms$ have a central SMBH, but this fraction rises to 100 percent above $10^8 \Ms$.\cite{Habouzit17}.

\subsubsection{Dynamical friction}

If the BH is not initially the central BH of the most massive progenitor halo, it depends on dynamical friction forces
\begin{equation}
    F_{\rm DF} = - \frac{4\pi (GM_{\rm BH})^2 \rho}{v^2} \ln \Lambda,
\end{equation}
to sink to the center, which occurs when matter accumulates in the wake of a massive object that creates a gravitational drag force.  Here $\rho$ is the mass density and $v$ is the BH velocity. The factor $\ln \Lambda = \ln(r_{\rm max} / r_{\rm min})$ is the Coulomb logarithm with $r_{\rm min}$ and $r_{\rm max}$ being the smallest and largest spatial scales that contribute to dynamical friction.

We can estimate an upper limit to the dynamical friction timescale $\tau_{\rm df}$ by calculating how long it takes for an object to lose all of its initial angular momentum as it experiences a force $F_{\rm DF}$.  The BH first follows its host halo with mass $M_{\rm host}$ as it falls into the main halo with mass $M_{\rm main}$.  Nearly all of the dynamical friction will be caused by the host halo during the galaxy merger.  The host  will then be tidally stripped and incorporated into the main halo, leaving the BH at some radius without its original host galaxy.  Now the BH has to sink to the center from dynamical friction caused by itself alone.  We can express the timescale in terms of the halo dynamical time $\tau_{\rm dyn} \equiv r_{\rm vir}/V_{\rm c}$ at the virial radius of the main halo,
\begin{equation}
    \frac{\tau_{\rm df}}{\tau_{\rm dyn}} = 1.17 \; \frac{f_{\rm df} \Theta_{\rm orb}}{\ln \Lambda} \; \frac{M_{\rm main}}{M_{\rm i}},
\end{equation}
where $f_{\rm df}$ is an adjustable parameter and $\Theta_{\rm orb}$ encodes information about the orbital energy and initial angular momentum.\cite{Binney87}.  Here $M_{\rm i} \in \{M_{\rm host}, M_{\rm BH}\}$ and depends on whether the BH is still embedded in its original host galaxy or not.  From this expression, one can see that light seeds will migrate more slowly to the halo centers than heavy seeds.

\subsubsection{Central black hole mergers}

Galaxy mergers, from first passage to complete coalescence, can take a substantial fraction (tens to hundreds of millions of years) of the Hubble time at $z > 10$.  After the galaxy merger completes, the two central BHs must sink into the potential well center through dynamical friction.  For heavy seeds, this is especially important because the ``close pair'' and ``dynamical heating'' scenarios have them forming in satellite halos that are falling into a more massive galaxy.  On the other hand, heavy seeds formed from streaming velocity effects would not have this problem.

The tight time constraints for the $z \sim 6$ quasars suggests that SMBHs grow mainly through accretion rather than mergers.\cite{Volonteri05}.  BH mergers also have the complication of the resulting BH receiving a ``kick'' (i.e. boost in linear momentum) that is usually on the order of 100\kms, higher than the escape velocity of low-mass halos.  These kicked BHs can escape the galaxy and take a substantial fraction of the Hubble time to return to the galaxy.\cite{Haiman04, Micic06}

% High-redshift galaxies tend to have high specific star formation rates ($\dot{M}_\star / M_\star \simeq 1 - 50 \unit{Gyr}^{-1})$, which indicate that they can double its stellar mass in as little as 20 million years.  However, there is a lack of low- and moderate-luminosity AGN at $z \ge 6$.  One plausible explanation, given high gas inflow rates in early galaxies, is that these central BHs are heavily obscured by dense and dusty gas.\cite{Pezzulli17}.

\section{Summary}

The properties of the first black holes in the Universe depend on their immediate environment throughout their lifetimes.  The variety of halo properties and large-scale environments leads to a diverse population of early BHs, ranging from stellar-mass to supermassive that may be either central to the host galaxy or roaming around its outskirts.

The properties of the host halo and molecular cloud control the formation of the stellar progenitor, its mass, multiplicity, and metallicity.  Also the cosmological environment plays a role in the LW radiation background intensity that controls \hh{} and Pop III star formation in minihalos.  Afterwards, the properties of the now-formed host galaxy and the nuclear star cluster affect the strength of BH accretion, ranging from inefficient to super-Eddington.

\begin{figure}[t]
    \centering
    \includegraphics[width=0.8\textwidth]{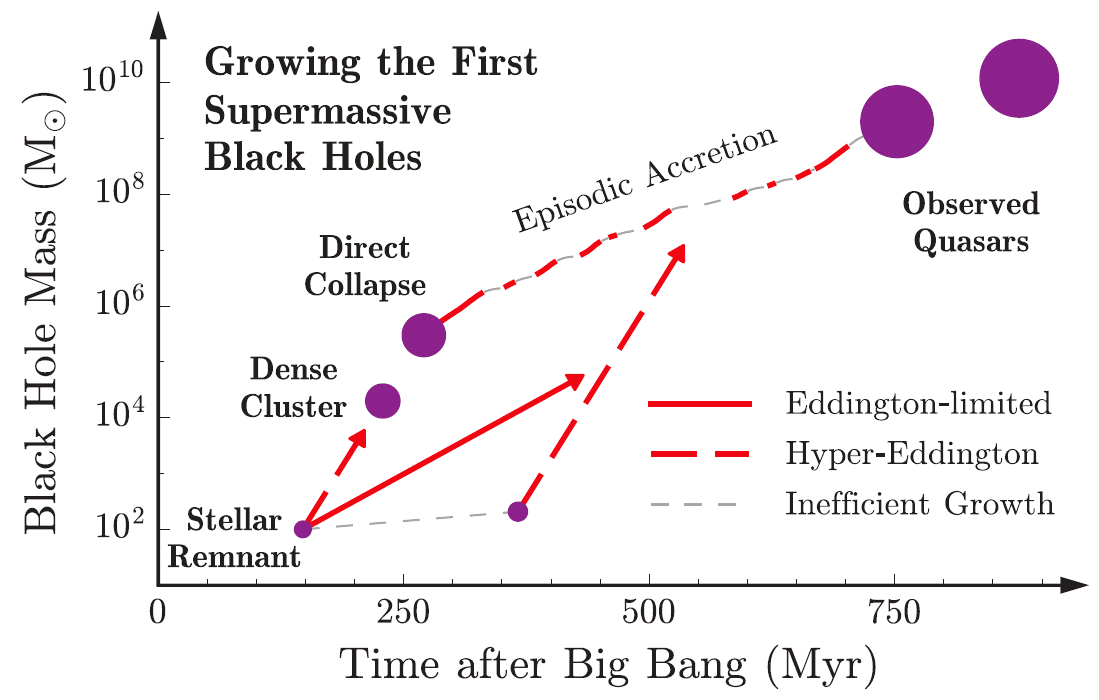}
    \caption{\label{fig:growth}\textbf{Possible growth histories of observed high-redshift quasars.}  Starting from three possible seeding mechanisms---light, intermediate, or heavy---the BHs can grow into the observed quasars at $z \sim 6$ through inefficient growth, Eddington-limited accretion, and/or hyper-accretion, or a combination of any of these processes. Adapted from Smith et al.\cite{Smith17_Review}}
\end{figure}

Figure \ref{fig:growth} shows a schematic view of possible growth histories of the observed $z \sim 6$ quasars, concisely depicting the seeding and growth of the first BHs in the universe.\cite{Smith17_Review}  It shows all three seeding scenarios (see Section \ref{ch9:sec:seeds}): a light BH seeded from a Pop III star, an intermediate mass BH seeded from collisions in a dense stellar cluster, a heavy BH seeded from a supermassive star.  The BHs will then experience a variety of relative growth rates, whether it be episodic, Eddington-limited, or super-Eddington, during its evolution to the present day.

\bibliographystyle{ws-rv-van}
\bibliography{jwise}

\begin{thebibliography}{116}
\providecommand{\natexlab}[1]{#1}
\providecommand{\url}[1]{\texttt{#1}}
\expandafter\ifx\csname urlstyle\endcsname\relax
  \providecommand{\doi}[1]{doi: #1}\else
  \providecommand{\doi}{doi: \begingroup \urlstyle{rm}\Url}\fi

\bibitem{Bromm13}
V.~{Bromm}, {Formation of the first stars}, \emph{Reports on Progress in
  Physics}. 76\penalty0 (11):\penalty0 112901  (Nov, 2013).
\newblock \doi{10.1088/0034-4885/76/11/112901}.

\bibitem{Oesch16}
P.~A. {Oesch}, G.~{Brammer}, P.~G. {van Dokkum} {\em et~al.}, {A Remarkably
  Luminous Galaxy at z=11.1 Measured with Hubble Space Telescope Grism
  Spectroscopy}, \emph{\apj}. 819:\penalty0 129  (Mar., 2016).
\newblock \doi{10.3847/0004-637X/819/2/129}.

\bibitem{Bouwens22}
R.~J. {Bouwens}, G.~D. {Illingworth}, P.~G. {van Dokkum} {\em et~al.}, {Sizes
  of Lensed Lower-luminosity z = 4-8 Galaxies from the Hubble Frontier Field
  Program}, \emph{\apj}. 927\penalty0 (1):\penalty0 81  (Mar., 2022).
\newblock \doi{10.3847/1538-4357/ac4791}.

\bibitem{Yang21}
J.~{Yang}, F.~{Wang}, X.~{Fan} {\em et~al.}, {Probing Early Supermassive Black
  Hole Growth and Quasar Evolution with Near-infrared Spectroscopy of 37
  Reionization-era Quasars at 6.3 < z {\ensuremath{\leq}} 7.64}, \emph{\apj}.
  923\penalty0 (2):\penalty0 262  (Dec., 2021).
\newblock \doi{10.3847/1538-4357/ac2b32}.

\bibitem{Wu15}
X.-B. {Wu}, F.~{Wang}, X.~{Fan} {\em et~al.}, {An ultraluminous quasar with a
  twelve-billion-solar-mass black hole at redshift 6.30}, \emph{\nat}. {\bf
  518}\penalty0 (7540), \penalty0 512--515  (Feb., 2015).
\newblock \doi{10.1038/nature14241}.

\bibitem{Banados18}
E.~{Ba{\~n}ados}, B.~P. {Venemans}, C.~{Mazzucchelli} {\em et~al.}, {An
  800-million-solar-mass black hole in a significantly neutral Universe at a
  redshift of 7.5}, \emph{\nat}. {\bf 553}\penalty0 (7689), \penalty0 473--476
  (Jan, 2018).
\newblock \doi{10.1038/nature25180}.

\bibitem{Fan06}
X.~{Fan}, C.~L. {Carilli} and B.~{Keating}, {Observational Constraints on
  Cosmic Reionization}, \emph{\araa}. {\bf 44}, \penalty0 415--462  (Sept.,
  2006).
\newblock \doi{10.1146/annurev.astro.44.051905.092514}.

\bibitem{Gultekin09}
K.~{G{\"u}ltekin}, D.~O. {Richstone}, K.~{Gebhardt} {\em et~al.}, {The
  M-{$\sigma$} and M-L Relations in Galactic Bulges, and Determinations of
  Their Intrinsic Scatter}, \emph{\apj}. {\bf 698}, \penalty0 198--221  (June,
  2009).
\newblock \doi{10.1088/0004-637X/698/1/198}.

\bibitem{McGreer18}
I.~D. {McGreer}, X.~{Fan}, L.~{Jiang} {\em et~al.}, {The Faint End of the z = 5
  Quasar Luminosity Function from the CFHTLS}, \emph{\aj}. 155\penalty0
  (3):\penalty0 131  (Mar, 2018).
\newblock \doi{10.3847/1538-3881/aaaab4}.

\bibitem{Somerville08}
R.~S. {Somerville}, P.~F. {Hopkins}, T.~J. {Cox} {\em et~al.}, {A semi-analytic
  model for the co-evolution of galaxies, black holes and active galactic
  nuclei}, \emph{\mnras}. {\bf 391}, \penalty0 481--506  (Dec., 2008).
\newblock \doi{10.1111/j.1365-2966.2008.13805.x}.

\bibitem{Willott10}
C.~J. {Willott}, P.~{Delorme}, C.~{Reyl{\'e}} {\em et~al.}, {The Canada-France
  High-z Quasar Survey: Nine New Quasars and the Luminosity Function at
  Redshift 6}, \emph{\aj}. {\bf 139}, \penalty0 906--918  (Mar., 2010).
\newblock \doi{10.1088/0004-6256/139/3/906}.

\bibitem{Volonteri17}
M.~{Volonteri}, A.~E. {Reines}, H.~{Atek} {\em et~al.}, {High-redshift Galaxies
  and Black Holes Detectable with the JWST: A Population Synthesis Model from
  Infrared to X-Rays}, \emph{\apj}. 849\penalty0 (2):\penalty0 155  (Nov.,
  2017).
\newblock \doi{10.3847/1538-4357/aa93f1}.

\bibitem{Woods19}
T.~E. {Woods}, B.~{Agarwal}, V.~{Bromm} {\em et~al.}, {Titans of the early
  Universe: The Prato statement on the origin of the first supermassive black
  holes}, \emph{\pasa}. 36:\penalty0 e027  (Aug., 2019).
\newblock \doi{10.1017/pasa.2019.14}.

\bibitem{Prole22}
L.~R. {Prole}, P.~C. {Clark}, R.~S. {Klessen} {\em et~al.},
  {Fragmentation-induced starvation in Population III star formation: a
  resolution study}, \emph{\mnras}. {\bf 510}\penalty0 (3), \penalty0
  4019--4030  (Mar., 2022).
\newblock \doi{10.1093/mnras/stab3697}.

\bibitem{Asplund09}
M.~{Asplund}, N.~{Grevesse}, A.~J. {Sauval} {\em et~al.}, {The Chemical
  Composition of the Sun}, \emph{\araa}. {\bf 47}\penalty0 (1), \penalty0
  481--522  (Sept., 2009).
\newblock \doi{10.1146/annurev.astro.46.060407.145222}.

\bibitem{Patridge67}
R.~B. {Partridge} and P.~J.~E. {Peebles}, {Are Young Galaxies Visible? II. The
  Integrated Background}, \emph{\apj}. {\bf 148}, \penalty0 377  (May, 1967).
\newblock \doi{10.1086/149161}.

\bibitem{Tumlinson10}
J.~{Tumlinson}, {Chemical Evolution in Hierarchical Models of Cosmic Structure.
  II. The Formation of the Milky Way Stellar Halo and the Distribution of the
  Oldest Stars}, \emph{\apj}. {\bf 708}, \penalty0 1398--1418  (Jan., 2010).
\newblock \doi{10.1088/0004-637X/708/2/1398}.

\bibitem{Caffau11}
E.~{Caffau}, P.~{Bonifacio}, P.~{Fran{\c{c}}ois} {\em et~al.}, {An extremely
  primitive star in the Galactic halo}, \emph{\nat}. {\bf 477}\penalty0 (7362),
  \penalty0 67--69  (Sept., 2011).
\newblock \doi{10.1038/nature10377}.

\bibitem{Hirano15}
S.~{Hirano}, T.~{Hosokawa}, N.~{Yoshida} {\em et~al.}, {Primordial star
  formation under the influence of far ultraviolet radiation: 1540 cosmological
  haloes and the stellar mass distribution}, \emph{\mnras}. {\bf 448},
  \penalty0 568--587  (Mar., 2015).
\newblock \doi{10.1093/mnras/stv044}.

\bibitem{Hosokawa16}
T.~{Hosokawa}, S.~{Hirano}, R.~{Kuiper} {\em et~al.}, {Formation of Massive
  Primordial Stars: Intermittent UV Feedback with Episodic Mass Accretion},
  \emph{\apj}. 824:\penalty0 119  (June, 2016).
\newblock \doi{10.3847/0004-637X/824/2/119}.

\bibitem{Ricotti01}
M.~{Ricotti}, N.~Y. {Gnedin} and J.~M. {Shull}, {Feedback from Galaxy
  Formation: Production and Photodissociation of Primordial H$_{2}$},
  \emph{\apj}. {\bf 560}, \penalty0 580--591  (Oct., 2001).
\newblock \doi{10.1086/323051}.

\bibitem{Field66}
G.~B. {Field}, W.~B. {Somerville} and K.~{Dressler}, {Hydrogen Molecules in
  Astronomy}, \emph{\araa}. {\bf 4}, \penalty0 207  (Jan., 1966).
\newblock \doi{10.1146/annurev.aa.04.090166.001231}.

\bibitem{Stecher67}
T.~P. {Stecher} and D.~A. {Williams}, {Photodestruction of Hydrogen Molecules
  in H I Regions}, \emph{\apjl}. {\bf 149}, \penalty0 L29  (July, 1967).
\newblock \doi{10.1086/180047}.

\bibitem{Wolcott12}
J.~{Wolcott-Green} and Z.~{Haiman}, {Feedback from the infrared background in
  the early Universe}, \emph{\mnras}. {\bf 425}, \penalty0 L51--L55  (Sept.,
  2012).
\newblock \doi{10.1111/j.1745-3933.2012.01298.x}.

\bibitem{Haiman97}
Z.~{Haiman}, M.~J. {Rees} and A.~{Loeb}, {Destruction of Molecular Hydrogen
  during Cosmological Reionization}, \emph{\apj}. {\bf 476}\penalty0 (2),
  \penalty0 458--463  (Feb., 1997).
\newblock \doi{10.1086/303647}.

\bibitem{Draine96}
B.~T. {Draine} and F.~{Bertoldi}, {Structure of Stationary Photodissociation
  Fronts}, \emph{\apj}. {\bf 468}, \penalty0 269--+  (Sept., 1996).
\newblock \doi{10.1086/177689}.

\bibitem{Wolcott11}
J.~{Wolcott-Green}, Z.~{Haiman} and G.~L. {Bryan}, {Photodissociation of
  H$_{2}$ in protogalaxies: modelling self-shielding in three-dimensional
  simulations}, \emph{\mnras}. p. 1673  (Oct., 2011).
\newblock \doi{10.1111/j.1365-2966.2011.19538.x}.

\bibitem{Wolcott19}
J.~{Wolcott-Green} and Z.~{Haiman}, {H$_{2}$ self-shielding with non-LTE
  rovibrational populations: implications for cooling in protogalaxies},
  \emph{\mnras}. {\bf 484}\penalty0 (2), \penalty0 2467--2473  (Apr., 2019).
\newblock \doi{10.1093/mnras/sty3280}.

\bibitem{Haiman96}
Z.~{Haiman}, A.~A. {Thoul} and A.~{Loeb}, {Cosmological Formation of Low-Mass
  Objects}, \emph{\apj}. {\bf 464}, \penalty0 523  (June, 1996).
\newblock \doi{10.1086/177343}.

\bibitem{Tegmark97}
M.~{Tegmark}, J.~{Silk}, M.~J. {Rees} {\em et~al.}, {How Small Were the First
  Cosmological Objects?}, \emph{\apj}. {\bf 474}, \penalty0 1--+  (Jan., 1997).
\newblock \doi{10.1086/303434}.

\bibitem{Visbal14}
E.~{Visbal}, Z.~{Haiman} and G.~L. {Bryan}, {Direct collapse black hole
  formation from synchronized pairs of atomic cooling haloes}, \emph{\mnras}.
  {\bf 445}, \penalty0 1056--1063  (Nov., 2014).
\newblock \doi{10.1093/mnras/stu1794}.

\bibitem{Tselia10}
D.~{Tseliakhovich} and C.~{Hirata}, {Relative velocity of dark matter and
  baryonic fluids and the formation of the first structures}, \emph{\prd}. {\bf
  82}\penalty0 (8), \penalty0 083520--+  (Oct., 2010).
\newblock \doi{10.1103/PhysRevD.82.083520}.

\bibitem{Tselia11}
D.~{Tseliakhovich}, R.~{Barkana} and C.~M. {Hirata}, {Suppression and spatial
  variation of early galaxies and minihaloes}, \emph{\mnras}. p. 1501  (Sept.,
  2011).
\newblock \doi{10.1111/j.1365-2966.2011.19541.x}.

\bibitem{Naoz11}
S.~{Naoz}, N.~{Yoshida} and R.~{Barkana}, {The non-linear evolution of baryonic
  overdensities in the early universe: initial conditions of numerical
  simulations}, \emph{\mnras}. {\bf 416}, \penalty0 232--241  (Sept., 2011).
\newblock \doi{10.1111/j.1365-2966.2011.19025.x}.

\bibitem{Kulkarni21}
M.~{Kulkarni}, E.~{Visbal} and G.~L. {Bryan}, {The Critical Dark Matter Halo
  Mass for Population III Star Formation: Dependence on Lyman-Werner Radiation,
  Baryon-dark Matter Streaming Velocity, and Redshift}, \emph{\apj}.
  917\penalty0 (1):\penalty0 40  (Aug., 2021).
\newblock \doi{10.3847/1538-4357/ac08a3}.

\bibitem{Schauer21}
A.~T.~P. {Schauer}, S.~C.~O. {Glover}, R.~S. {Klessen} {\em et~al.}, {The
  influence of streaming velocities and Lyman-Werner radiation on the formation
  of the first stars}, \emph{\mnras}. {\bf 507}\penalty0 (2), \penalty0
  1775--1787  (Oct., 2021).
\newblock \doi{10.1093/mnras/stab1953}.

\bibitem{Skinner20}
D.~{Skinner} and J.~H. {Wise}, {Cradles of the first stars: self-shielding,
  halo masses, and multiplicity}, \emph{\mnras}. {\bf 492}\penalty0 (3),
  \penalty0 4386--4397  (Mar., 2020).
\newblock \doi{10.1093/mnras/staa139}.

\bibitem{Yoshida03}
N.~{Yoshida}, T.~{Abel}, L.~{Hernquist} {\em et~al.}, {Simulations of Early
  Structure Formation: Primordial Gas Clouds}, \emph{\apj}. {\bf 592},
  \penalty0 645--663  (Aug., 2003).
\newblock \doi{10.1086/375810}.

\bibitem{Omukai01}
K.~{Omukai}, {Primordial Star Formation under Far-Ultraviolet Radiation},
  \emph{\apj}. {\bf 546}, \penalty0 635--651  (Jan., 2001).
\newblock \doi{10.1086/318296}.

\bibitem{Wolcott17}
J.~{Wolcott-Green}, Z.~{Haiman} and G.~L. {Bryan}, {Beyond J$_{crit}$: a
  critical curve for suppression of H$_{2}$-cooling in protogalaxies},
  \emph{\mnras}. {\bf 469}\penalty0 (3), \penalty0 3329--3336  (Aug., 2017).
\newblock \doi{10.1093/mnras/stx167}.

\bibitem{Sutherland93}
R.~S. {Sutherland} and M.~A. {Dopita}, {Cooling Functions for Low-Density
  Astrophysical Plasmas}, \emph{\apjs}. {\bf 88}, \penalty0 253  (Sept., 1993).
\newblock \doi{10.1086/191823}.

\bibitem{Fernandez14}
R.~{Fernandez}, G.~L. {Bryan}, Z.~{Haiman} {\em et~al.}, {H$_{2}$ suppression
  with shocking inflows: testing a pathway for supermassive black hole
  formation}, \emph{\mnras}. {\bf 439}\penalty0 (4), \penalty0 3798--3807
  (Apr., 2014).
\newblock \doi{10.1093/mnras/stu230}.

\bibitem{Turk09}
M.~J. {Turk}, T.~{Abel} and B.~{O'Shea}, {The Formation of Population III
  Binaries from Cosmological Initial Conditions}, \emph{Science}. {\bf 325},
  \penalty0 601--  (July, 2009).
\newblock \doi{10.1126/science.1173540}.

\bibitem{Greif12}
T.~H. {Greif}, V.~{Bromm}, P.~C. {Clark} {\em et~al.}, {Formation and evolution
  of primordial protostellar systems}, \emph{\mnras}. {\bf 424}\penalty0 (1),
  \penalty0 399--415  (July, 2012).
\newblock \doi{10.1111/j.1365-2966.2012.21212.x}.

\bibitem{Hirano17_P3}
S.~{Hirano} and V.~{Bromm}, {Formation and survival of Population III stellar
  systems}, \emph{\mnras}. {\bf 470}\penalty0 (1), \penalty0 898--914  (Sept.,
  2017).
\newblock \doi{10.1093/mnras/stx1220}.

\bibitem{Susa19}
H.~{Susa}, {Merge or Survive: Number of Population III Stars per Minihalo},
  \emph{\apj}. 877\penalty0 (2):\penalty0 99  (June, 2019).
\newblock \doi{10.3847/1538-4357/ab1b6f}.

\bibitem{Kroupa02}
P.~{Kroupa}, {The Initial Mass Function of Stars: Evidence for Uniformity in
  Variable Systems}, \emph{Science}. {\bf 295}, \penalty0 82--91  (Jan., 2002).
\newblock \doi{10.1126/science.1067524}.

\bibitem{Heger03}
A.~{Heger}, C.~L. {Fryer}, S.~E. {Woosley} {\em et~al.}, {How Massive Single
  Stars End Their Life}, \emph{\apj}. {\bf 591}, \penalty0 288--300  (July,
  2003).
\newblock \doi{10.1086/375341}.

\bibitem{Fowler64}
W.~A. {Fowler}, {Massive Stars, Relativistic Polytropes, and Gravitational
  Radiation}, \emph{Reviews of Modern Physics}. {\bf 36}\penalty0 (2),
  \penalty0 545--554  (Apr., 1964).
\newblock \doi{10.1103/RevModPhys.36.545}.

\bibitem{Hosokawa11}
T.~{Hosokawa}, K.~{Omukai}, N.~{Yoshida} {\em et~al.}, {Protostellar Feedback
  Halts the Growth of the First Stars in the Universe}, \emph{Science}. {\bf
  334}, \penalty0 1250--  (Dec., 2011).
\newblock \doi{10.1126/science.1207433}.

\bibitem{Krumholz14}
M.~R. {Krumholz}, {The big problems in star formation: The star formation rate,
  stellar clustering, and the initial mass function}, \emph{\physrep}. {\bf
  539}, \penalty0 49--134  (June, 2014).
\newblock \doi{10.1016/j.physrep.2014.02.001}.

\bibitem{Loeb94}
A.~{Loeb} and F.~A. {Rasio}, {Collapse of primordial gas clouds and the
  formation of quasar black holes}, \emph{\apj}. {\bf 432}, \penalty0 52--61
  (Sept., 1994).
\newblock \doi{10.1086/174548}.

\bibitem{Dijkstra08}
M.~{Dijkstra}, Z.~{Haiman}, A.~{Mesinger} {\em et~al.}, {Fluctuations in the
  high-redshift Lyman-Werner background: close halo pairs as the origin of
  supermassive black holes}, \emph{\mnras}. {\bf 391}, \penalty0 1961--1972
  (Dec., 2008).
\newblock \doi{10.1111/j.1365-2966.2008.14031.x}.

\bibitem{Regan16}
J.~A. {Regan}, P.~H. {Johansson} and J.~H. {Wise}, {Forming Super-Massive Black
  Hole Seeds under the Influence of a Nearby Anisotropic Multi-Frequency
  Source}, \emph{\mnras}  (Apr., 2016).
\newblock \doi{10.1093/mnras/stw899}.

\bibitem{Wise19}
J.~H. {Wise}, J.~A. {Regan}, B.~W. {O'Shea} {\em et~al.}, {Formation of massive
  black holes in rapidly growing pre-galactic gas clouds}, \emph{\nat}. {\bf
  566}\penalty0 (7742), \penalty0 85--88  (Jan, 2019).
\newblock \doi{10.1038/s41586-019-0873-4}.

\bibitem{Hirano17}
S.~{Hirano}, T.~{Hosokawa}, N.~{Yoshida} {\em et~al.}, {Supersonic gas streams
  enhance the formation of massive black holes in the early universe},
  \emph{Science}. {\bf 357}, \penalty0 1375--1378  (Sept., 2017).
\newblock \doi{10.1126/science.aai9119}.

\bibitem{Milo09_IMBH}
M.~{Milosavljevi{\'c}}, S.~M. {Couch} and V.~{Bromm}, {Accretion Onto
  Intermediate-Mass Black Holes in Dense Protogalactic Clouds}, \emph{\apjl}.
  {\bf 696}, \penalty0 L146--L149  (May, 2009).
\newblock \doi{10.1088/0004-637X/696/2/L146}.

\bibitem{Park11_IMBH1}
K.~{Park} and M.~{Ricotti}, {Accretion onto Intermediate-mass Black Holes
  Regulated by Radiative Feedback. I. Parametric Study for Spherically
  Symmetric Accretion}, \emph{\apj}. 739:\penalty0 2  (Sept., 2011).
\newblock \doi{10.1088/0004-637X/739/1/2}.

\bibitem{Begelman78}
M.~C. {Begelman} and M.~J. {Rees}, {The fate of dense stellar systems},
  \emph{\mnras}. {\bf 185}, \penalty0 847--860  (Dec, 1978).
\newblock \doi{10.1093/mnras/185.4.847}.

\bibitem{kitayama04}
T.~{Kitayama}, N.~{Yoshida}, H.~{Susa} {\em et~al.}, {The Structure and
  Evolution of Early Cosmological H II Regions}, \emph{\apj}. {\bf 613},
  \penalty0 631--645  (Oct., 2004).
\newblock \doi{10.1086/423313}.

\bibitem{Whalen04}
D.~{Whalen}, T.~{Abel} and M.~L. {Norman}, {Radiation Hydrodynamic Evolution of
  Primordial H II Regions}, \emph{\apj}. {\bf 610}, \penalty0 14--22  (July,
  2004).
\newblock \doi{10.1086/421548}.

\bibitem{Abel07}
T.~{Abel}, J.~H. {Wise} and G.~L. {Bryan}, {The H II Region of a Primordial
  Star}, \emph{\apjl}. {\bf 659}, \penalty0 L87--L90  (Apr., 2007).
\newblock \doi{10.1086/516820}.

\bibitem{Alvarez09}
M.~A. {Alvarez}, J.~H. {Wise} and T.~{Abel}, {Accretion onto the First
  Stellar-Mass Black Holes}, \emph{\apjl}. {\bf 701}\penalty0 (2), \penalty0
  L133--L137  (Aug., 2009).
\newblock \doi{10.1088/0004-637X/701/2/L133}.

\bibitem{Jeon12}
M.~{Jeon}, A.~H. {Pawlik}, T.~H. {Greif} {\em et~al.}, {The First Galaxies:
  Assembly with Black Hole Feedback}, \emph{\apj}. 754:\penalty0 34  (July,
  2012).
\newblock \doi{10.1088/0004-637X/754/1/34}.

\bibitem{Smith18}
B.~D. {Smith}, J.~A. {Regan}, T.~P. {Downes} {\em et~al.}, {The growth of black
  holes from Population III remnants in the Renaissance simulations},
  \emph{\mnras}. {\bf 480}\penalty0 (3), \penalty0 3762--3773  (Nov, 2018).
\newblock \doi{10.1093/mnras/sty2103}.

\bibitem{Omukai08}
K.~{Omukai}, R.~{Schneider} and Z.~{Haiman}, {Can Supermassive Black Holes Form
  in Metal-enriched High-Redshift Protogalaxies?}, \emph{\apj}. {\bf
  686}\penalty0 (2), \penalty0 801--814  (Oct., 2008).
\newblock \doi{10.1086/591636}.

\bibitem{Devecchi09}
B.~{Devecchi} and M.~{Volonteri}, {Formation of the First Nuclear Clusters and
  Massive Black Holes at High Redshift}, \emph{\apj}. {\bf 694}, \penalty0
  302--313  (Mar., 2009).
\newblock \doi{10.1088/0004-637X/694/1/302}.

\bibitem{Katz15}
H.~{Katz}, D.~{Sijacki} and M.~G. {Haehnelt}, {Seeding high-redshift QSOs by
  collisional runaway in primordial star clusters}, \emph{\mnras}. {\bf 451},
  \penalty0 2352--2369  (Aug., 2015).
\newblock \doi{10.1093/mnras/stv1048}.

\bibitem{Tagawa20}
H.~{Tagawa}, Z.~{Haiman} and B.~{Kocsis}, {Making a Supermassive Star by
  Stellar Bombardment}, \emph{\apj}. 892\penalty0 (1):\penalty0 36  (Mar.,
  2020).
\newblock \doi{10.3847/1538-4357/ab7922}.

\bibitem{Schleicher22}
D.~R.~G. {Schleicher}, B.~{Reinoso}, M.~{Latif} {\em et~al.}, {Origin of
  supermassive black holes in massive metal-poor protoclusters}, \emph{\mnras}.
  {\bf 512}\penalty0 (4), \penalty0 6192--6200  (June, 2022).
\newblock \doi{10.1093/mnras/stac926}.

\bibitem{Davies11}
M.~B. {Davies}, M.~C. {Miller} and J.~M. {Bellovary}, {Supermassive Black Hole
  Formation Via Gas Accretion in Nuclear Stellar Clusters}, \emph{\apjl}.
  740\penalty0 (2):\penalty0 L42  (Oct., 2011).
\newblock \doi{10.1088/2041-8205/740/2/L42}.

\bibitem{Alexander14}
T.~{Alexander} and P.~{Natarajan}, {Rapid growth of seed black holes in the
  early universe by supra-exponential accretion}, \emph{Science}. {\bf 345},
  \penalty0 1330--1333  (Sept., 2014).
\newblock \doi{10.1126/science.1251053}.

\bibitem{Sakurai19}
Y.~{Sakurai}, N.~{Yoshida} and M.~S. {Fujii}, {Growth of intermediate mass
  black holes by tidal disruption events in the first star clusters},
  \emph{\mnras}. {\bf 484}\penalty0 (4), \penalty0 4665--4677  (Apr, 2019).
\newblock \doi{10.1093/mnras/stz315}.

\bibitem{Park16}
K.~{Park}, M.~{Ricotti}, P.~{Natarajan} {\em et~al.}, {Bulge-driven Fueling of
  Seed Black Holes}, \emph{\apj}. 818:\penalty0 184  (Feb., 2016).
\newblock \doi{10.3847/0004-637X/818/2/184}.

\bibitem{Begelman06}
M.~C. {Begelman}, M.~{Volonteri} and M.~J. {Rees}, {Formation of supermassive
  black holes by direct collapse in pre-galactic haloes}, \emph{\mnras}. {\bf
  370}, \penalty0 289--298  (July, 2006).
\newblock \doi{10.1111/j.1365-2966.2006.10467.x}.

\bibitem{Begelman08}
M.~C. {Begelman}, E.~M. {Rossi} and P.~J. {Armitage}, {Quasi-stars: accreting
  black holes inside massive envelopes}, \emph{\mnras}. {\bf 387}, \penalty0
  1649--1659  (July, 2008).
\newblock \doi{10.1111/j.1365-2966.2008.13344.x}.

\bibitem{Johnson13_DCBH}
J.~L. {Johnson}, D.~J. {Whalen}, H.~{Li} {\em et~al.}, {Supermassive Seeds for
  Supermassive Black Holes}, \emph{\apj}. 771:\penalty0 116  (July, 2013).
\newblock \doi{10.1088/0004-637X/771/2/116}.

\bibitem{Montero12}
P.~J. {Montero}, H.-T. {Janka} and E.~{M{\"u}ller}, {Relativistic Collapse and
  Explosion of Rotating Supermassive Stars with Thermonuclear Effects},
  \emph{\apj}. 749:\penalty0 37  (Apr., 2012).
\newblock \doi{10.1088/0004-637X/749/1/37}.

\bibitem{Whalen13_SMS}
D.~J. {Whalen}, W.~{Even}, J.~{Smidt} {\em et~al.}, {Supermassive Population
  III Supernovae and the Birth of the First Quasars}, \emph{\apj}.
  778:\penalty0 17  (Nov., 2013).
\newblock \doi{10.1088/0004-637X/778/1/17}.

\bibitem{Matsumoto15}
T.~{Matsumoto}, D.~{Nakauchi}, K.~{Ioka} {\em et~al.}, {Can Direct Collapse
  Black Holes Launch Gamma-Ray Bursts and Grow to Supermassive Black Holes?},
  \emph{\apj}. 810:\penalty0 64  (Sept., 2015).
\newblock \doi{10.1088/0004-637X/810/1/64}.

\bibitem{Agarwal13}
B.~{Agarwal}, A.~J. {Davis}, S.~{Khochfar} {\em et~al.}, {Unravelling obese
  black holes in the first galaxies}, \emph{\mnras}. {\bf 432}\penalty0 (4),
  \penalty0 3438--3444  (July, 2013).
\newblock \doi{10.1093/mnras/stt696}.

\bibitem{Barrow18}
K.~S.~S. {Barrow}, J.~H. {Wise}, A.~{Aykutalp} {\em et~al.}, {First light - II.
  Emission line extinction, population III stars, and X-ray binaries},
  \emph{\mnras}. {\bf 474}, \penalty0 2617--2634  (Feb., 2018).
\newblock \doi{10.1093/mnras/stx2973}.

\bibitem{Regan20_VMS}
J.~A. {Regan}, J.~H. {Wise}, T.~E. {Woods} {\em et~al.}, {The Formation of Very
  Massive Stars in Early Galaxies and Implications for Intermediate Mass Black
  Holes}, \emph{The Open Journal of Astrophysics}. 3\penalty0 (1):\penalty0 15
  (Dec., 2020).
\newblock \doi{10.21105/astro.2008.08090}.

\bibitem{Carr21}
B.~{Carr}, K.~{Kohri}, Y.~{Sendouda} {\em et~al.}, {Constraints on primordial
  black holes}, \emph{Reports on Progress in Physics}. 84\penalty0
  (11):\penalty0 116902  (Nov., 2021).
\newblock \doi{10.1088/1361-6633/ac1e31}.

\bibitem{Poulin17}
V.~{Poulin}, P.~D. {Serpico}, F.~{Calore} {\em et~al.}, {CMB bounds on
  disk-accreting massive primordial black holes}, \emph{\prd}. 96\penalty0
  (8):\penalty0 083524  (Oct., 2017).
\newblock \doi{10.1103/PhysRevD.96.083524}.

\bibitem{PadmanabhanBook}
T.~{Padmanabhan}, \emph{{Structure Formation in the Universe}}  (1993).

\bibitem{Luo11}
B.~{Luo}, W.~N. {Brandt}, Y.~Q. {Xue} {\em et~al.}, {Revealing a Population of
  Heavily Obscured Active Galactic Nuclei at z {\ensuremath{\approx}} 0.5-1 in
  the Chandra Deep Field-South}, \emph{\apj}. 740\penalty0 (1):\penalty0 37
  (Oct., 2011).
\newblock \doi{10.1088/0004-637X/740/1/37}.

\bibitem{Wang16}
F.~{Wang}, X.-B. {Wu}, X.~{Fan} {\em et~al.}, {A Survey of Luminous
  High-redshift Quasars with SDSS and WISE. I. Target Selection and Optical
  Spectroscopy}, \emph{\apj}. 819\penalty0 (1):\penalty0 24  (Mar., 2016).
\newblock \doi{10.3847/0004-637X/819/1/24}.

\bibitem{Decarli18}
R.~{Decarli}, F.~{Walter}, B.~P. {Venemans} {\em et~al.}, {An ALMA [C II]
  Survey of 27 Quasars at z > 5.94}, \emph{\apj}. 854\penalty0 (2):\penalty0 97
   (Feb., 2018).
\newblock \doi{10.3847/1538-4357/aaa5aa}.

\bibitem{Shimasaku19}
K.~{Shimasaku} and T.~{Izumi}, {Black versus Dark: Rapid Growth of Supermassive
  Black Holes in Dark Matter Halos at z {\ensuremath{\sim}} 6}, \emph{\apjl}.
  872\penalty0 (2):\penalty0 L29  (Feb., 2019).
\newblock \doi{10.3847/2041-8213/ab053f}.

\bibitem{Volonteri11}
M.~{Volonteri} and D.~P. {Stark}, {Assessing the redshift evolution of massive
  black holes and their hosts}, \emph{\mnras}. {\bf 417}\penalty0 (3),
  \penalty0 2085--2093  (Nov., 2011).
\newblock \doi{10.1111/j.1365-2966.2011.19391.x}.

\bibitem{Lupi19}
A.~{Lupi}, M.~{Volonteri}, R.~{Decarli} {\em et~al.}, {High-redshift quasars
  and their host galaxies - I. Kinematical and dynamical properties and their
  tracers}, \emph{\mnras}. {\bf 488}\penalty0 (3), \penalty0 4004--4022
  (Sept., 2019).
\newblock \doi{10.1093/mnras/stz1959}.

\bibitem{Wise07}
J.~H. {Wise} and T.~{Abel}, {Resolving the Formation of Protogalaxies. I.
  Virialization}, \emph{\apj}. {\bf 665}, \penalty0 899--910  (Aug., 2007).
\newblock \doi{10.1086/520036}.

\bibitem{Greif08}
T.~H. {Greif}, J.~L. {Johnson}, R.~S. {Klessen} {\em et~al.}, {The first
  galaxies: assembly, cooling and the onset of turbulence}, \emph{\mnras}. {\bf
  387}, \penalty0 1021--1036  (July, 2008).
\newblock \doi{10.1111/j.1365-2966.2008.13326.x}.

\bibitem{Regan09}
J.~A. {Regan} and M.~G. {Haehnelt}, {Pathways to massive black holes and
  compact star clusters in pre-galactic dark matter haloes with virial
  temperatures {$>$}\~{}10000K}, \emph{\mnras}. {\bf 396}, \penalty0 343--353
  (June, 2009).
\newblock \doi{10.1111/j.1365-2966.2009.14579.x}.

\bibitem{Lodato06}
G.~{Lodato} and P.~{Natarajan}, {Supermassive black hole formation during the
  assembly of pre-galactic discs}, \emph{\mnras}. {\bf 371}\penalty0 (4),
  \penalty0 1813--1823  (Oct., 2006).
\newblock \doi{10.1111/j.1365-2966.2006.10801.x}.

\bibitem{Lodato07}
G.~{Lodato} and P.~{Natarajan}, {The mass function of high-redshift seed black
  holes}, \emph{\mnras}. {\bf 377}\penalty0 (1), \penalty0 L64--L68  (May,
  2007).
\newblock \doi{10.1111/j.1745-3933.2007.00304.x}.

\bibitem{Bondi44}
H.~{Bondi} and F.~{Hoyle}, {On the mechanism of accretion by stars},
  \emph{\mnras}. {\bf 104}, \penalty0 273  (1944).
\newblock \doi{10.1093/mnras/104.5.273}.

\bibitem{Bondi47}
H.~{Bondi}, F.~{Hoyle} and R.~A. {Lyttleton}, {On the structure of the solar
  corona and chromosphere}, \emph{\mnras}. {\bf 107}, \penalty0 184  (Jan.,
  1947).
\newblock \doi{10.1093/mnras/107.2.184}.

\bibitem{Bondi52}
H.~{Bondi}, {On spherically symmetrical accretion}, \emph{\mnras}. {\bf 112},
  \penalty0 195  (1952).
\newblock \doi{10.1093/mnras/112.2.195}.

\bibitem{Balbus91}
S.~A. {Balbus} and J.~F. {Hawley}, {A powerful local shear instability in
  weakly magnetized disks. I - Linear analysis. II - Nonlinear evolution},
  \emph{\apj}. {\bf 376}, \penalty0 214--233  (July, 1991).
\newblock \doi{10.1086/170270}.

\bibitem{Power11}
C.~{Power}, S.~{Nayakshin} and A.~{King}, {The accretion disc particle method
  for simulations of black hole feeding and feedback}, \emph{\mnras}. {\bf
  412}, \penalty0 269--276  (Mar., 2011).
\newblock \doi{10.1111/j.1365-2966.2010.17901.x}.

\bibitem{Abramowicz88}
M.~A. {Abramowicz}, B.~{Czerny}, J.~P. {Lasota} {\em et~al.}, {Slim accretion
  disks}, \emph{\apj}. {\bf 332}, \penalty0 646--658  (Sept., 1988).
\newblock \doi{10.1086/166683}.

\bibitem{Novak12}
G.~S. {Novak}, J.~P. {Ostriker} and L.~{Ciotti}, {Radiative transfer and
  radiative driving of outflows in active galactic nuclei and starbursts},
  \emph{\mnras}. {\bf 427}, \penalty0 2734--2756  (Dec., 2012).
\newblock \doi{10.1111/j.1365-2966.2012.21844.x}.

\bibitem{Jiang14}
Y.-F. {Jiang}, J.~M. {Stone} and S.~W. {Davis}, {A Global Three-dimensional
  Radiation Magneto-hydrodynamic Simulation of Super-Eddington Accretion
  Disks}, \emph{\apj}. 796:\penalty0 106  (Dec., 2014).
\newblock \doi{10.1088/0004-637X/796/2/106}.

\bibitem{VanBorm13}
C.~{Van Borm} and M.~{Spaans}, {The influence of magnetic fields, turbulence,
  and UV radiation on the formation of supermassive black holes}, \emph{\aap}.
  553:\penalty0 L9  (May, 2013).
\newblock \doi{10.1051/0004-6361/201321590}.

\bibitem{Begelman79}
M.~C. {Begelman}, {Can a spherically accreting black hole radiate very near the
  Eddington limit?}, \emph{\mnras}. {\bf 187}, \penalty0 237--251  (Apr.,
  1979).
\newblock \doi{10.1093/mnras/187.2.237}.

\bibitem{Inayoshi16}
K.~{Inayoshi}, Z.~{Haiman} and J.~P. {Ostriker}, {Hyper-Eddington accretion
  flows on to massive black holes}, \emph{\mnras}. {\bf 459}\penalty0 (4),
  \penalty0 3738--3755  (Jul, 2016).
\newblock \doi{10.1093/mnras/stw836}.

\bibitem{Pacucci17}
F.~{Pacucci}, P.~{Natarajan}, M.~{Volonteri} {\em et~al.}, {Conditions for
  Optimal Growth of Black Hole Seeds}, \emph{\apjl}. 850\penalty0 (2):\penalty0
  L42  (Dec., 2017).
\newblock \doi{10.3847/2041-8213/aa9aea}.

\bibitem{volonteri09}
M.~{Volonteri} and N.~Y. {Gnedin}, {Relative Role of Stars and Quasars in
  Cosmic Reionization}, \emph{\apj}. {\bf 703}, \penalty0 2113--2117  (Oct.,
  2009).
\newblock \doi{10.1088/0004-637X/703/2/2113}.

\bibitem{Habouzit17}
M.~{Habouzit}, M.~{Volonteri} and Y.~{Dubois}, {Blossoms from black hole seeds:
  properties and early growth regulated by supernova feedback}, \emph{\mnras}.
  {\bf 468}, \penalty0 3935--3948  (July, 2017).
\newblock \doi{10.1093/mnras/stx666}.

\bibitem{Binney87}
J.~{Binney} and S.~{Tremaine}, \emph{{Galactic dynamics}}  (1987).

\bibitem{Volonteri05}
M.~{Volonteri} and M.~J. {Rees}, {Rapid Growth of High-Redshift Black Holes},
  \emph{\apj}. {\bf 633}, \penalty0 624--629  (Nov., 2005).
\newblock \doi{10.1086/466521}.

\bibitem{Haiman04}
Z.~{Haiman}, {Constraints from Gravitational Recoil on the Growth of
  Supermassive Black Holes at High Redshift}, \emph{\apj}. {\bf 613}\penalty0
  (1), \penalty0 36--40  (Sept., 2004).
\newblock \doi{10.1086/422910}.

\bibitem{Micic06}
M.~{Micic}, T.~{Abel} and S.~{Sigurdsson}, {The role of primordial kicks on
  black hole merger rates}, \emph{\mnras}. {\bf 372}, \penalty0 1540--1548
  (Nov., 2006).
\newblock \doi{10.1111/j.1365-2966.2006.11013.x}.

\bibitem{Smith17_Review}
A.~{Smith}, V.~{Bromm} and A.~{Loeb}, {The first supermassive black holes},
  \emph{Astronomy and Geophysics}. {\bf 58}\penalty0 (3), \penalty0 3.22--3.26
  (June, 2017).
\newblock \doi{10.1093/astrogeo/atx099}.

\end{thebibliography}

%\printindex[aindx]                 % to print author index
%\printindex                         % to print subject index
\end{document}